\documentclass[12pt]{article}
\pdfoutput=1                          

\usepackage{amsmath}
\usepackage{slashed}
\usepackage{amssymb}
\usepackage{epsfig}
\usepackage{graphicx, bm}
\usepackage{multirow}
\usepackage{color}
\usepackage[normal]{subfigure}
\usepackage{rotating}
\usepackage{hyperref}
\usepackage[margin=1in]{geometry}
\usepackage[table]{xcolor}
\usepackage{enumitem}
\usepackage[utf8x]{inputenc}
\usepackage[compress,numbers,sort]{natbib}
\usepackage{authblk}
\usepackage{colortbl}
\usepackage{pdflscape}
\usepackage{xspace}

\usepackage{journals}

\definecolor{LightCyan}{rgb}{0.88,1,1}
\definecolor{piggypink}{rgb}{0.99, 0.87, 0.9}

\usepackage{ucs}
\usepackage{amsmath}
\usepackage{amsfonts}
\usepackage{amssymb}
\usepackage{eucal}               
\usepackage{mathrsfs}          
\usepackage[sans]{dsfont}     
\usepackage[Symbol]{upgreek}  

\usepackage{array}             
\usepackage{booktabs}       
\usepackage{colortbl}          
\usepackage{multirow}
\usepackage{longtable}

\usepackage{cleveref}

\usepackage{fancyhdr}

\usepackage[normalem]{ulem}             
\usepackage{cancel}           

\usepackage{slashed}
\usepackage{simplewick}     
\usepackage{feynmf}           

\usepackage{color}
\definecolor{grigio}{cmyk}{0,0,0,0.1}
\definecolor{rosa}{cmyk}{0,0.1,0.1,0.02}
\definecolor{rosino}{cmyk}{0,0.05,0.05,0.02}
\definecolor{rosas}{cmyk}{0,0.3,0.25,0.05}
\definecolor{celeste}{cmyk}{0.1,0,0,0.02}
\definecolor{giallino}{cmyk}{0,0,0.1,0.02}
\definecolor{rosso}{cmyk}{0,1,1,0.4}
\definecolor{rossos}{cmyk}{0,1,1,0.55}
\definecolor{rossoc}{cmyk}{0,1,1,0.2}
\definecolor{blu}{cmyk}{1,1,0,0.3}
\definecolor{blus}{cmyk}{1,1,0,0.5}
\definecolor{bluc}{cmyk}{1,1,0,0.1}
\definecolor{blucc}{cmyk}{0.7,0.5,0,0}
\definecolor{viola}{cmyk}{0,1,0,0.6}
\definecolor{viola2}{cmyk}{0,1,0.2,0.6}
\definecolor{verde}{cmyk}{0.92,0,0.59,0.25}
\definecolor{verdec}{cmyk}{0.92,0,0.59,0.15}
\definecolor{verdes}{cmyk}{0.92,0,0.59,0.4}
\definecolor{verdino}{cmyk}{0.12,0,0.09,0.02}
\definecolor{giallo}{cmyk}{0,0,1,0}
\definecolor{gialloverde}{cmyk}{0.44,0,0.74,0}
\definecolor{Titolo}{rgb}{0.752941176,0.576470588,0.992156863}
\definecolor{altro}{rgb}{0.094117647,0.650980392,0.643137255}
\definecolor{Peanuts}{rgb}{0.2, 0.4, 0.6}
\definecolor{Pean1}{rgb}{0.6, 0.8, 0.4}
\definecolor{BHO}{rgb}{0.2, 0.8, 1}
\definecolor{Daria}{rgb}{0, 0.9412, 0}
\definecolor{UniPi}{rgb}{0.2549, 0.4627, 0.6275}
\definecolor{UniPidue}{rgb}{0.3216, 0.5804, 0.7882}
\definecolor{rossoCP3}{cmyk}{0,.88,.77,.40}
\definecolor{verdeCP3}{rgb}{0.09765625, 0.57421875, 0.1015625}
\definecolor{bluCP3}{rgb}{0, 0.23, 0.67}
\newcommand{\mDM}{m_\text{DM}}
\newcommand{\sigmav}{\langle \sigma v \rangle}
\newcommand{\ud}{\text{d}}

\newcommand{\beq}{\begin{equation}}
\newcommand{\eeq}{\end{equation}}

\def\LHC{{\sc LHC}}

\def\FERMI{{\sc Fermi}} 
\def\HESS{{\sc H.E.S.S.}}
\def\MAGIC{{\sc MAGIC}}
\def\VERITAS{{\sc VERITAS}}
\def\CTA{{\sc CTA}}

\def\Sag{{\tt  Sagittarius}}

\def\Car{{\tt  Carina}}
\def\Dra{{\tt  Draco}}
\def\For{{\tt  Fornax}}
\def\Scu{{\tt  Sculptor}}
\def\UMi{{\tt  Ursa Minor}}

\def\Boo{{\tt  Bootes~I}}
\def\Com{{\tt  Coma Berenices}}
\def\Seg{{\tt  Segue~I}}
\def\UMaI{{\tt  Ursa Major~I}}
\def\UMaII{{\tt  Ursa Major~II}}

\ifx\pdfoutput\undefined
\usepackage[dvips,bookmarks]{hyperref}    
\else
\usepackage{hyperref}    
\fi
\hypersetup{colorlinks, bookmarksopen, bookmarksnumbered,
citecolor=verdeCP3, linkcolor=bluCP3, pdfstartview=FitH, urlcolor=rossoCP3}

\numberwithin{equation}{section}

\begin{document}

\begin{flushright}
{\footnotesize
{\sc IRFU-15-36}
}
\end{flushright}
\color{black}

\begin{center}

{\Huge\bf Prospects for annihilating  Dark Matter   towards  Milky Way's dwarf galaxies by the Cherenkov \\[2mm]  Telescope Array}

\medskip
\bigskip\color{black}\vspace{0.6cm}

{
{\large\bf Valentin Lefranc}$\, ^{a}$,
{\large\bf Gary A. Mamon}$\, ^{b}$,
{\large\bf Paolo Panci}$\, ^{b}$
}
\\[7mm]
{\it $^a$ \href{http://irfu.cea.fr/}{IRFU/DSM}, CEA Saclay, F-91191 Gif-sur-Yvette, France} \\ [3mm]
{\it $^b$ \href{http://www.iap.fr}{Institut d'Astrophysique de Paris}, UMR
7095: (CNRS \& UPMC, Sorbonne Universit\'es), \\ 98 bis Boulevard Arago, Paris F-75014, France}\\[2mm]
\end{center}

\bigskip

\centerline{\large\bf Abstract}
\begin{quote}

We derive the  Cherenkov Telescope Array (\CTA) sensitivity to  dark matter (DM) annihilation in several primary channels, over a broad range of DM masses. These sensitivities are estimated when \CTA\ is pointed towards a large sample of Milky Way's dwarf spheroidal galaxies (dSphs) with promising $J$-factors and small statistical uncertainties.  This analysis neglects systematic uncertainties, which we estimate at the level of at least 1 dex. We also present sensitivities on the annihilation cross section from a combined analysis of 4 dSphs. We assess the \CTA\ sensitivity by: $i)$ using, for each dSph, a recent determination of the $J$-factor and its statistical error; $ii)$ considering the most up-to-date cosmic ray background; and $iii)$ including both spatial and spectral terms in the likelihood analysis. We find that a joint spectral and spatial analysis improves the \CTA\ sensitivity, in particular for primary channels with sharp features in the $\gamma$-ray energy spectrum and for dSphs with steep $J$-factor profiles, as deduced from the internal kinematics. The greatest sensitivities are obtained for observations of \UMi\ among the classical dSphs and of \UMaII\ for ultra-faint dSphs. 

\end{quote}

\newpage
\tableofcontents


\section{Introduction}
\label{Sec:Intro}
The Large Hadron Collider (\LHC) has now completed the Standard Model (SM) by revealing the presence of the Higgs boson. Nevertheless, so far, the \LHC\ has also found no clear evidence for new physics at TeV energy range, casting doubts on the \emph{naturalness}~\cite{'tHooft:1980xb} of the ElectroWeak (EW) scale and in turn on the very special role in elementary particle physics of such energy. On the other hand, the argument dubbed as the \emph{WIMP miracle} (where WIMPs are Weak Interacting Massive Particles), suggests that the TeV energy range is still interesting in Dark Matter (DM) terms, independently on whether it is relevant or not for the \emph{naturalness} problem. Indeed, a particle with a mass in the TeV range, interacting weekly with the ordinary matter, can naturally provide the observed DM abundance thanks to the \emph{thermal freeze-out} process.  Multiplets of the SM charged under the EW gauge group are, by definition, the prototype of WIMP DM and they arise as  candidates in motivated beyond SM constructions (e.g.~the supersymmetric Wino~\cite{Giudice:1998xp, Giudice:2004tc, ArkaniHamed:2004fb} and the candidates of the Minimal DM framework~\cite{Cirelli:2005uq, DelNobile:2015bqo}). Hence, betting on the possibility of a pure WIMP nature of DM, indirect searches of secondary stable SM products of DM annihilations turn out to be very promising (see e.g.~Refs.~\cite{Bergstrom:2000pn, Bertone:2004pz, Cirelli:2012tf} for a review of indirect detection of DM annihilation through $\gamma$-ray observations). 

\smallskip
More specifically, the current and upcoming ground-based telescopes are designed to identify the DM properties in the mass range of hundreds of GeV up to few tens of TeV, by detecting the Very High Energy (VHE, $E \gtrsim$ 100 GeV) $\gamma$-rays due to annihilating DM particles coming from different astrophysical objects (e.g.~the inner part of the Milky Way's halo, nearby Milky Way's dwarf spheroidal (dSph) satellites, and clusters of galaxies).

The inner halo of the Milky Way is often thought to be the most promising
laboratory to search for VHE $\gamma$-rays from annihilating DM, thanks to
its proximity to the Earth and the possibility for the DM distribution to
exhibit a high density cusp. However, the Galactic Center (GC) is also a very
crowded region for VHE $\gamma$-ray searches, because a very high
astrophysical background is expected. For example, the \HESS\
experiment~\cite{hess} detected strong VHE
sources~\cite{Aharonian:2004wa,Aharonian:2005br,Aharonian:2009zk} and diffuse
emission~\cite{Aharonian:2006au} extending along the Galactic Plane in the
inner 300 pc of the GC. At lower energies, besides the detection of sources
in the GC region~\cite{Acero:2015hja}, the \FERMI\ satellite~\cite{fermi}
revealed a Galactic diffuse emission~\cite{FermiLAT:2012aa} that appears to
be unrelated to DM and hence acts as an overwhelming source of
confusion. Indeed, the abundance of astrophysical sources and the complexity
of the GC region, make it very challenging to separate the diffuse emission
by annihilating/decaying DM from the astrophysical sources.  In spite of that, several authors have reported the detection of a $\gamma$-ray signal from the inner part of the GC~\cite{Hooper:2010mq, Abazajian:2012pn, Gordon:2013vta, Calore:2014xka, TheFermi-LAT:2015kwa}. Although its  spectrum and morphology are compatible with those expected from annihilating DM particles~\cite{Daylan:2014rsa, Calore:2014nla}, one should always  bear in mind that there might be alternative astrophysical explanations for such excess~\cite{Abazajian:2010zy, Petrovic:2014uda, Cholis:2015dea, Gaggero:2015nsa, Carlson:2016iis}.

On the other hand, dSphs are probably the cleanest laboratories to look for DM in VHE $\gamma$-rays, thanks to their high DM content and reduced stellar emission foreground.  Indeed, their stellar dynamics indicate that dSphs are among the most DM dominated objects in the Universe~(see Ref.~\cite{Courteau+14} and references therein). Moreover, classical and ultra-faint dSph galaxies show little signs of either recent stellar formation activity (see Refs.~\cite{Weisz+11} and \cite{Brown+12}, respectively) or gas acting as target material for cosmic rays propagation in the interstellar medium (see Ref.~\cite{Mateo98} and references therein). In particular, the dSphs orbiting the Milky Way at typically 100 kpc or less from the GC are compelling targets for indirect DM searches with $\gamma$-rays~\cite{2013PhR5311S}.
   
\smallskip  
More specifically, over the last past decade, the \FERMI\  satellite~\cite{fermi} and the Imaging Atmospheric (or Air) \v{C}erenkov Telescopes (IACTs), such as \HESS~\cite{hess}, \MAGIC~\cite{magic} and \VERITAS~\cite{veritas}, have carried out various observational campaigns towards nearby dSphs, and no significant $\gamma$-ray emission has been detected so far.\footnote{A 3.2$\,\sigma$ detection of a $\gamma$-ray excess at energies between 2 to 5 GeV has been reported from an analysis of pass~7 data from \FERMI\ \cite{Geringer-Sameth:2015lua, Hooper:2015ula}, but the significance of this detection is only 1.65$\,\sigma$ significant with the pass~8 data \cite{Drlica-Wagner:2015xua}.} At low energies ($E \lesssim 300$ GeV), the recent bounds on the annihilation cross section, coming from a 6-year \FERMI\ data analysis of 15 dSphs, challenge the thermal freeze out value up to DM masses of roughly 100 GeV for the hadronic channels (e.g.~40 to 200 GeV at 68\%
containment and  20 to 300 GeV at 95\% containment for the DM DM $\rightarrow b\bar b$ primary mode). At higher energies, the constraints imposed by \HESS~\cite{Aharonian:2008dm, Aharonian:2007km, Abramowski:2010aa, Abramowski:2014tra}, \MAGIC~\cite{Aliu:2008ny, Albert:2007xg,Aleksic:2013xea}, and \VERITAS~\cite{Acciari:2010pja,2012PhRvD85f2001A}, complement the bounds of \FERMI\ for heavy DM candidates. In particular, the latest constraints from \HESS~\cite{Abramowski:2014tra}, coming from the observation of a subset of 4 dSphs, plus the \Sag\ dwarf, rule out DM annihilation cross sections of the order of $10^{-23}\,\rm cm^3\,s^{-1}$ for the hadronic channels, in the DM mass window of $1\rightarrow20$ TeV.

The forthcoming Cherenkov Telescope Array (\CTA), will bring up to a factor of ten improvement in terms of flux sensitivity compared to currently operating IACTs~\cite{2011ExA32193A}, with a factor two to three in angular and energy resolutions. This will be fundamental for two main reasons: $i)$ for well-motivated multi-TeV WIMP DM candidates (e.g.~supersymmetric Wino DM, Minimal DM candidates), the annihilation cross sections into SM particles could receive a significant boost at low velocity due to Sommerfeld corrections by orders of magnitudes~\cite{Hisano:2006nn, Cirelli:2007xd, Cirelli:2015bda, Garcia-Cely:2015dda, DelNobile:2015bqo}. Therefore, a gain in sensitivity with respect to currently operating IACTs, can be sufficient to probe almost the entire parameter space of many well motivated WIMP models at the TeV-scale; $ii)$ given the angular resolution of \CTA, dSph galaxies will be no longer point-like objects. This is important because one can then implement a spatial analysis of the likelihood, which will improve the \CTA\ sensitivity.

In this paper, we provide a new assessment of the \CTA\ sensitivity to DM annihilations towards nearby Milky Way's Satellite galaxies. More specifically, 
\begin{itemize}
\item[$\diamond$] for each dSph, we use a recent determination of the $J$-factor profile and its statistical error, coming from analyses of stellar-kinematic data (see e.g.~\cite{Geringer-Sameth:2014yza, Bonnivard:2015xpq, Bonnivard:2015tta}).

\item[$\diamond$] as already done in Ref.~\cite{Lefranc:2015pza}, we consider the most up-to-date determination of the irreducible background coming from a full \CTA\ Monte Carlo simulation.

\item[$\diamond$] we implement three different statistical approaches in order to quantitatively show how the \CTA\ sensitivity is improved when jointly considering the different spectral and spatial behaviors of the DM signal  compared to the irreducible background.   

\item[$\diamond$] we include a Gaussian term in the likelihood in order to measure  the impact of the $J$-factor statistical uncertainties on the \CTA\ sensitivity.

\item[$\diamond$]  we  thoroughly describe the sources of systematic uncertainties in the derivations of the $J$-factors. 
\end{itemize}

The remaining of this article is organized as follows. In Sec.~\ref{Sec:CTA}, we briefly review the \CTA\ performances in the context of dSph galaxies observations; in Sec.~\ref{Sec:DMdSphs}, we shortly summarize the properties of the $\gamma$-ray fluxes from DM annihilations,  our  selection of  dSph galaxies and the \emph{Regions of Interest} (RoIs) considered in our analysis. We also review the expected number of DM signal  and background  events in the \CTA\ array, together with a discussion on the possible sources of systematic uncertainties on the $J$-factor determinations; Sec.~\ref{Sec:AnaMet} presents our analysis methodology. In particular, the implementation of both the spectral and spatial analysis in the \CTA\ likelihood considering the statistical uncertainties on the $J$-factor computations as well; in Sec.~\ref{Sec:Results}, we display the results of our analyses, and we conclude in Sec.~\ref{Sec:Summary}.

\section{CTA performances}
\label{Sec:CTA}
\CTA\ is the next-generation observatory for ground-based $\gamma$-ray
astronomy. It is envisaged as a two-site array, one in each hemisphere (Roque de los Muchachos Observatory in La Palma in the Canary Islands, Spain and European Southern Observatory Cerro Paranal in Chile).  According to Ref.~\cite{performance}, \CTA\  will surpass the overall performances of the currently operating IACTs: the field of view (FoV) will be wider (around  7.5$^\circ$ for the middle size telescopes), the angular resolution will two to three times better (3 arcmin at 68\% containment radius at 1 TeV energies,  corresponding to 15 arcmin at full-width half-maximum (FWHM)),  an order of magnitude improvement on flux sensitivity, an energy resolution better than 10\% between 0.5 and 10 TeV, and a lower threshold of a few tens of GeV (see also Ref.~\cite{Acharya:2013sxa}).  Although the final layout of the array is not yet finalized, detailed Monte Carlo performance studies have been conducted on several candidate arrays in order to characterize their instrument response functions in terms of background rejection, flux sensitivity, energy and angular resolutions~\cite{2013APh43171B}.  In the present article, we consider the instrument response function of the candidate array~I and we make use of the effective area, residual background rate, angular and energy resolutions provided in Ref.~\cite{2013APh43171B}.

As for the Galactic Center (GC) studies, observations of dSph galaxies with IACTs usually employ two regions of the sky, one dubbed as the ON region for signal extraction, and one called OFF region for background measurements. The ON and OFF regions are chosen to be fairly nearby in the sky and the search for a $\gamma$-ray excess proceeds with a test statistics method between the two regions. As it will become clearer later on,  observations of the Milky Way's dSph galaxies allow for simultaneous measurements of $\gamma$-ray counts in
the ON and OFF regions. This alleviates systematic uncertainties on the determination of possible excesses that may arise from different observational and instrumental conditions. Furthermore, the ON-OFF technique for signal and background measurements with \CTA\ is very well suited for dSphs. Indeed, regardless of the DM profiles, the  \CTA\ FoV is always much larger than the angular size of the dSph, which always enables a significant gradient between the ON and OFF regions. This is in contrast to the GC studies, where the ON-OFF technique is only reliable for a  cuspy GC DM profile~\cite{Lefranc:2015pza}.

In the present study, we optimize the above mentioned ON-OFF method towards nearby dSph galaxies by carrying out a full likelihood analysis that takes into account the spectral and spatial distributions of the $\gamma$-rays arising from DM annihilation. Our analysis takes full advantage of the radial distribution  of the signal with respect to the background by using the available information well beyond the spatial point-spread-function of the instrument. In what follows, we will consider a uniform exposure in the RoIs.

\section{Dark matter annihilation in dwarf galaxies}
\label{Sec:DMdSphs}

\subsection{Analysis Setup}
\label{Sec:AnalysisSetup}
The $\gamma$-ray flux from self-annihilation of DM particles of mass $\mDM$ in a dSph galaxy measured at angle $\theta'$ between the direction to the center of the object and the line of sight (l.o.s.) is 
\beq\label{flux}
\frac{\ud \Phi}{\ud \Omega \ud E} = \frac1{8\pi \mDM^2}\sum_f
\sigmav_f \frac{\ud N_f}{\ud E}(E) \frac{\ud
  J(\theta')}{\ud \Omega} \ , 
\eeq
where  $\sigmav_f$ and $\ud N_f/\ud E$  are respectively the thermally-averaged velocity-weighted annihilation cross section and the energy spectrum of photons per one annihilation in the channel with final state $f$, respectively. Eq.~\eqref{flux} shows that the $\gamma$-ray flux is the product of a particle physics term (including cross section, DM particle mass and energy spectrum) and an astrophysical term, called the \emph{$J$-factor}, involving the l.o.s. integral of the square of the DM mass density, 
\beq\label{Jdef}
\frac{\ud J(\theta')}{\ud \Omega}= \int_{\rm l.o.s.} \rho^2 \, 
\ud \ell \ ,
\eeq
where $\ell$ is the l.o.s. coordinate.

\smallskip
We compute the $\gamma$-ray energy spectrum using the tools given in Ref.~\cite{Cirelli:2010xx}, and we only consider the \emph{prompt} contribution to the $\gamma$-ray flux. In principle, at low energies, the contribution from inverse Compton secondary emission must be also taken into account. Indeed, this is particularly important for the determination of the \CTA\ sensitivity in leptonic annihilation channels, as demonstrated in Ref.~\cite{Lefranc:2015pza}. Nevertheless, since the photon energy density in dSph is much lower than in GC, we expect a negligible contribution from inverse Compton emission. Indeed, the mean free path of $\sim$ 10 TeV electrons is in general much larger that the size of the dSph galaxies.

\subsubsection{Choice of the dSphs}
\label{Sec:dSphs}

\begin{table}[!t]
\tabcolsep 8pt
\center
\scriptsize{\begin{tabular}{l|c|c|c|c|c}
Target & Hemisphere &Distance [kpc] & Angular size  [deg] & Number of RoIs
& log$_{10} (J_{\rm all} \ [\rm GeV^2\,cm^{-5}$]) \\
\hline
\hline
\Car\ &S& 101 & 1.26 & 4 &17.84$^{+0.08}_{-0.08}$\\
\Dra\ & N&\ \,82 & 1.30 & 6 &18.89$^{+0.14}_{-0.14}$\\
\For\ & S&138 & 2.61 & 3 &17.78$^{+0.13}_{-0.08}$\\
\Scu\ & S&\ \,79 & 1.94 & 3 &18.45$^{+0.07}_{-0.06}$\\
\UMi\ &N&\ \,66& 1.37 & 2 &18.89$^{+0.30}_{-0.30}$\\
\hline
\hline
\Boo\ &N &\ \,66 & 0.47 & 4&18.20$^{+0.40}_{-0.36}$\\
\Com\ &N&\ \,44 & 0.31 & 3 &19.02$^{+0.37}_{-0.40}$\\
\Seg\ &N&\ \,23 & 0.35 & 3&19.33$^{+0.32}_{-0.34}$\\
\UMaI\ &N&\ \,97& 0.43 & 4&17.86$^{+0.56}_{-0.33}$\\
\UMaII\ &N&\ \,30& 0.53 & 4& 19.36$^{+0.42}_{-0.41}$\\
\hline
\hline
\end{tabular}}
\caption{ \small \label{tab:dSphs} List of dSph galaxies considered in our analysis with their relevant properties. The second, third and fourth columns show the celestial hemisphere where the object is located,  the distance  from the observer, and  the angular size if the dSph (angular distance of its last member star), respectively. In the last two columns, we report the optimized number of 6 arcmin wide RoIs to detect a DM annihilation signal and the logarithm of the $J$-factor, summed over all the RoIs, and its statistical error.} 
\end{table}

As stated in the Introduction, dSph galaxies of the Milky Way are promising targets for indirect DM searches via VHE $\gamma$-rays, thanks to their high DM content, their closeness and relatively low stellar emission foreground. In 2005, there were  9 luminous ``classical'' dSph galaxies identified as Milky Way satellites. In the last decade, the Sloan Digital Sky Survey (SDSS), has extended the dwarf galaxy regime to extremely low sizes and luminosities, making possible the discovery of new faint surface-brightness objects, usually referred as ``ultra-faint'' dSph galaxies~\cite{Belokurov+07}.  This larger sample of Milky Way dSph galaxies and the prospects for the detection of new ones with future surveys like PanSTARRS~\cite{PanSTARRS,Kaiser:2002zz}, Dark Energy Survey (DES)~\cite{DES,Flaugher:2004vg} and Large Synoptic Survey Telescope (LSST)~\cite{LSST,Tyson:2002nh,Hargis14b}, are generating excitement and activity in this field.

On the other hand, given the scarcity of stellar tracers, the determination
of the $J$-factor in a given dSph, is affected by quite large uncertainties
that are the subject of a long ongoing debate (see,
e.g.,~\cite{Evans:2003sc,Strigari:2007at,Essig:2009jx,Charbonnier:2011ft,GeringerSameth:2011iw,Martinez:2013els}). For
a stacked sample of dSph galaxies, the \FERMI\ collaboration finds small
statistical uncertainty at the level of 10\% to
40\%~\cite{Ackermann:2015zua}. For the individual classical dSph galaxy, a
recent work quotes values from 17\% to
124\%~\cite{Geringer-Sameth:2014yza}. Another recent
study~\cite{Bonnivard:2015xpq} finds larger uncertainties, which can however
be different from one case to another. In particular, they point out that the
astrophysical factors of \Seg\ might be highly uncertain due to few kinematic
data and probable stellar contamination. In the past month, many studies
appear in the literature providing $J$-factors of several new
dSphs~\cite{Hayashi:2016kcy, Ullio:2016kvy, Evans:2016xwx, Genina:2016kzg, Drlica-Wagner:2015xua}. In particular, for the first time, Ref.~\cite{Hayashi:2016kcy} evaluates the $J$-factors for 24 dSphs taking into account non-spherical dark halos. Finally, in case of exotic scenarios in which dSph galaxies host black holes, the $J$-factor could be deeply affected (from a factor of a few up to $10^6$~\cite{Gonzalez-Morales:2014eaa}). In view of this rather unclear situation, Sec.~\ref{Sec:SysUncertainties}, is devoted to a critical discussion about the sources of systematic uncertainties that come with the computations of the $J$-factor.

\smallskip
We base our analysis  on Ref.~\cite{Geringer-Sameth:2014yza}, which tabulates
values and uncertainties of the astrophysical $J$-factor as a function of
$\theta$  for several dSphs. We therefore do not include any new dSphs, in
particular those detected from the DES survey \cite{Drlica-Wagner:2015xua},
because of the lack of resolved $J$ factors.
We select from the list of dSphs in  \cite{Geringer-Sameth:2014yza}, 
those with the most promising $J$-factors and
smallest statistical uncertainties (less than 10\%). As a result, we choose 5 classical dSph galaxies (\Car, \Dra, \For, \Scu, \UMi), and 5 ultra-faint ones (\Boo, \Com, \Seg, \UMaI, \UMaII). A subset of these dSphs were already considered by the currently operating IACTs in various observational campaigns. In particular, \Seg\ was a target of primary interest in terms of DM searches~\cite{Aleksic:2011jx,2012PhRvD85f2001A,Aleksic:2013xea} before the dedicated study in  Ref.~\cite{Bonnivard:2015xpq} came out. We still consider this candidate in our analysis, keeping however well in mind that the value of the $J$-factor quoted in Ref.~\cite{Geringer-Sameth:2014yza} might be largely overestimated (at least a factor $\mathcal O(10)$ as shown in Fig.~7 of Ref.~\cite{Bonnivard:2015xpq}). 

The list of dSph galaxies considered in our analysis is given in the first column of Tab.~\ref{tab:dSphs}. We also report, in the second, third and fourth columns, the best location of \CTA\ for observing such objects, the distances  from the observer and angular sizes in the sky, respectively~\cite{Geringer-Sameth:2014yza, Drlica-Wagner:2015xua}. 

\subsubsection{Choice of the regions of interest}
\label{Sec:RoIs}
In our spatial analysis, we are interested in non-overlapping regions. Hence, the simplest choice is to consider RoIs corresponding to annuli centered on the dSph with inner angular radius $(i-1)\,\Delta \theta$ and outer angular radius $i\,\Delta\theta$, where the
1st RoI is circle of radius $\Delta\theta$ on the sky. As often done in astronomy,  the width $\Delta \theta$ of the annuli is chosen to be 2.5 times smaller than the typical PSF FWHM, i.e. $\Delta\theta=6\,\rm arcmin$.

For a circular aperture of angular radius $\theta$ centered on a dSph, the $J$-factor is easily obtained from Eq.~\eqref{Jdef} 
\begin{equation}
J_{\rm ap}(\theta) =2\pi \int_0^\theta \ud \theta'  \sin\theta'\, \frac{\ud J(\theta')}{\ud \Omega} \ , 
\label{Jap}
\end{equation}
and therefore in the $i$th RoI it  writes
\begin{equation}
J_i = 
J_{\rm ap}(i \Delta\theta)-J_{\rm ap}((i\!-\!1)\Delta\theta) =2\pi  \int_{(i\!-\!1)\Delta\theta}^{i \Delta\theta} \ud \theta'  \sin\theta'\, \frac{\ud J(\theta')}{\ud \Omega} \ , \qquad \mbox{where } i>1 \ .
\label{Ji}
\end{equation}
According to Eq.~\eqref{flux}, the $\gamma$-ray flux for the $i$th RoI will then be 
\begin{equation}
{\ud \Phi_i\over \ud E} = 
\frac{J_i}{8\pi \mDM^2}\sum_f
\sigmav_f \frac{\ud N_f}{\ud E}(E) \ . 
\label{eq:fluxannulus}
\end{equation}

We interpolate the values and uncertainties of $\log_{10} J_i$ from the values and uncertainties of $\log_{10} J_{\rm ap}(\theta)$ computed in Ref.~\cite{Geringer-Sameth:2014yza} and tabulated (in ASCII form) in logarithmic bins of $\theta$ in~\cite{jfactordata} to our linear grid of angular radii $i\,\Delta\theta = 6', 12', 18'$, ... Once the uncertainties on $\log_{10} J(\theta)$ are interpolated to our linear grid, the $1\sigma$  uncertainty  on $J_i$, hereafter called $\sigma_i$, is obtained by subtracting the corresponding uncertainties on $J_{\rm ap}(i\Delta\theta)$ and $J_{\rm ap}((i-1)\Delta\theta)$. Such a procedure does not introduce a systematic if the  correlation matrix  in our RoIs is diagonal.

The number of RoIs considered for each dSph depends on the DM distribution of the object. It is set on a case-by-case basis depending on the spatial behavior of $J_i$ in order to obtain the optimal sensitivity for each dSph considered in the study. The optimization is done in terms of signal-to-noise ratio where the signal and the isotropic irreducible background (see Sec.~\ref{Sec:Bkgevents}) are proportional to $J_i$ and $\Delta \Omega_i = 4\pi\left[\sin^2(i \Delta\theta/2)-\sin^2((i\!-\!1)\Delta\theta/2)\right]$, respectively.  More specifically, we first derive the \CTA\ sensitivity in the first RoI (cone with angular radius of 6 arcmin). Then, step by step, we add RoIs with increasing aperture until the \CTA\ sensitivity becomes saturated. This procedure allows us to determine the maximal aperture $\theta_{\rm max}$ of the last annulus in a given dSph galaxy and in turn the number of RoIs within such an angle. In the last two columns of Tab.~\ref{tab:dSphs}, we report the optimized number of RoIs to observe a DM annihilation signal together with  the logarithmic value of the $J$-factor summed over all the RoIs ($\log_{10} J_{\rm all}$) and its $\pm$1$\sigma$ statistical error.

The optimized number of RoIs in dSphs that feature cuspy DM density profiles is smaller than those with a smooth ones. For example, one can compare the number of RoIs derived  for \Scu\ and \Dra, which respectively have steep and shallow $\gamma$-ray surface brightness profiles (respectively red squares and black circles in the right hand panel of Fig.~\ref{fig:1}):  the \CTA\ sensitivity is saturated after the 3rd RoI for \Scu, but only for the 6th RoI for \Dra. Hence,  we expect that dSph galaxies with smooth DM distributions can profit more from a spatial study due to the higher number of RoIs. However,  the \CTA\ sensitivity can be boosted in a relevant way towards dSph galaxies with cuspy DM profiles, because the DM signal is concentrated in the very inner regions
of these galaxies. Therefore, a signal-to-noise ratio optimization gets the maximal benefit. In Sec.~\ref{Sec:Results}, we show an explicit example of the projected \CTA\ sensitivity towards two dSphs with almost the same $J$-factors (\Dra\ and \UMi), despite completely different DM distributions.

\subsection{Dark matter signals for CTA}
\label{Sec:DMevents}
\begin{figure}[!t]
\centering
\includegraphics[width=.495\textwidth]{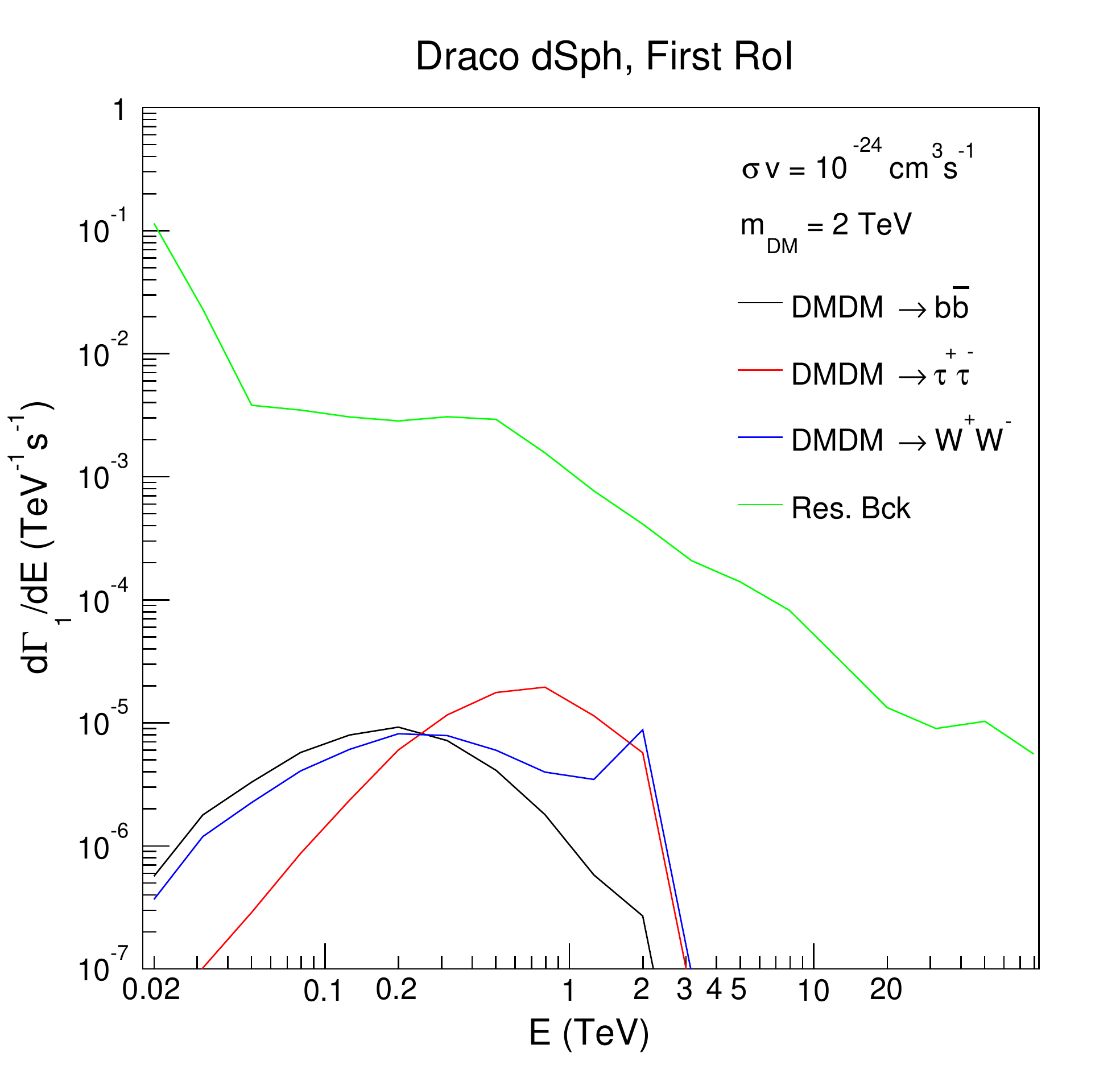}
\includegraphics[width=.495\textwidth]{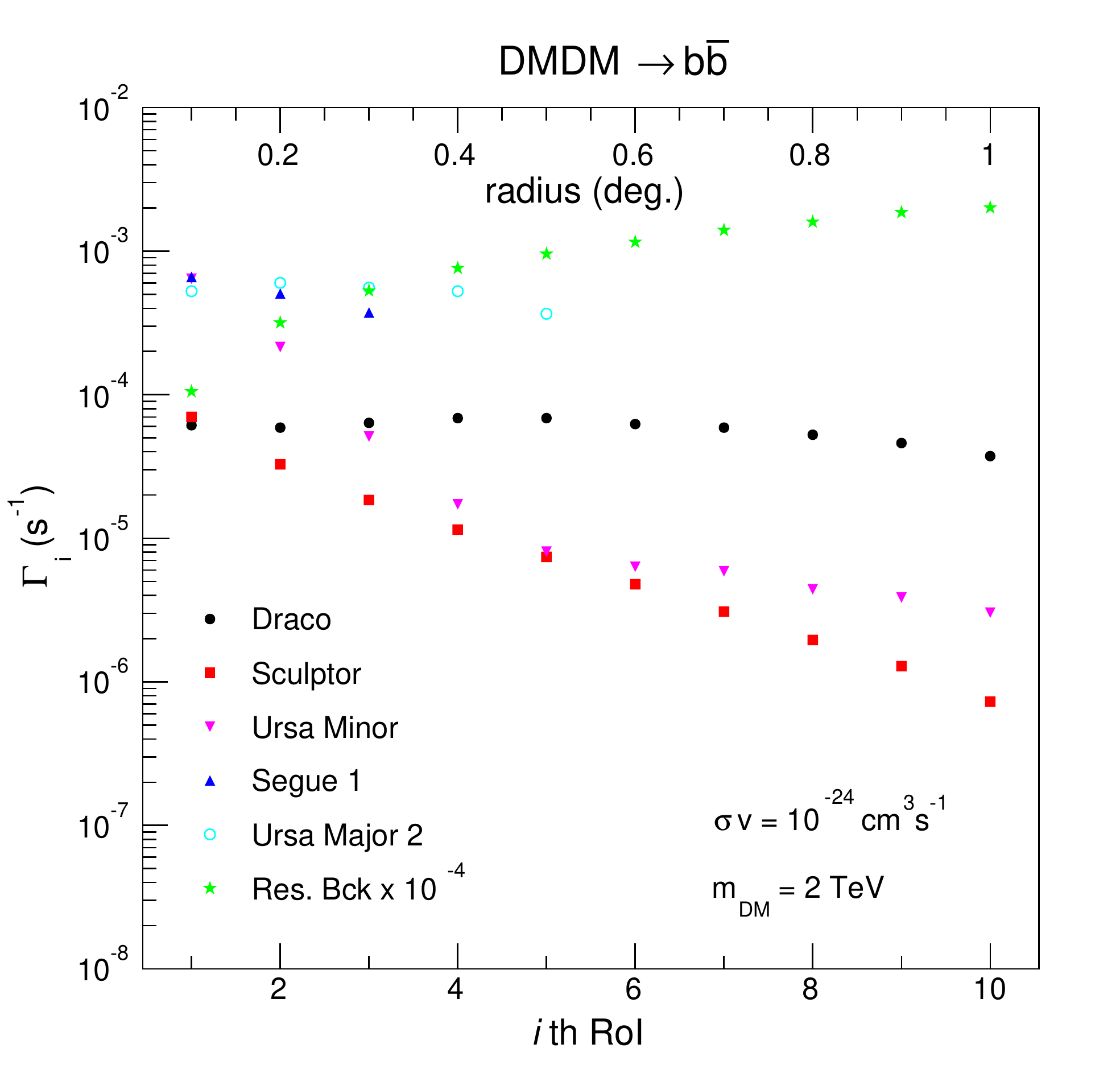}
\caption{\small {\it Left panel}: Predicted differential count rate spectra in the first RoI towards \Dra, for 2 TeV DM candidates annihilating with $\langle \sigma v \rangle = 10^{-24}\,\rm cm^3\,s^{-1}$ into $b \bar b$ (black solid line), $\tau^+\tau^-$ (red solid line) and $W^+W^-$ (blue solid line) primary channels. The most up-to-date residual background estimation coming from a \CTA\ Monte Carlo simulation is also shown (green solid line). {\it Right panel}: Predicted total rate summed over all energy bins above 30 GeV in the $i$th RoI, i.e., as a function of the outer angular radius of the $i$th RoI (top $x$-axis), considering 2 TeV DM candidates annihilating with $\langle \sigma v \rangle = 10^{-24}\,\rm cm^3\,s^{-1}$ into $ b \bar b$. The spatial dependence of the total rate is presented for three classical dSph galaxies (\Dra\ (black-filled dots), \Scu\ (red-filled squares) and \UMi\ (magenta-filled triangles)) and two ultra-faint ones  (\Seg\ (blue-filled triangles) and \UMaII\ (cyan circles)). The spatial variation of the irreducible background total rate multiplied by $10^{-4}$ is also presented for comparison (green-filled stars). See Sec.~\ref{Sec:DMdSphs} for further details. }
\label{fig:1}
\end{figure}

Given a certain observation time $\Delta t$, the number of observed $\gamma$-ray events in the \CTA\ array is computed by convolving the right hand side of Eq.~(\ref{eq:fluxannulus}) with the energy-dependent effective area for photon $\mathcal A_{\rm eff}(E)$ and the Gaussian energy response function. For each dSph, in the $i$th RoI and $j$th energy bin, the signal counts are
\begin{equation}
\begin{split}
\label{eq:countrate}
&N_{ij}^{\rm S}=\Delta t \int_{\Delta E_j} \hspace{-.15cm} \ud E 
\, \frac{\ud \Gamma_{ i}^{\rm S}}{\ud E}  \qquad \hbox{where}  \\ 
& \frac{\ud \Gamma_{ i}^{\rm S}}{\ud E}=
\int_{-\infty}^{+\infty} \hspace{-.15cm} \ud E' \, \frac{\ud \Phi_i}{ \ud
  E}(E') \, \mathcal A_{\rm eff}(E') \,  
\frac{\exp\left\{-(E\!-\!E')^2/\left[2\sigma^2(E')\right]\right\}}{\sqrt{2\pi\sigma^2(E')}} \ ,
\end{split}
\end{equation} 
where $\sigma(E) = \delta_{\rm res}(E) / \sqrt{8 \ln(2)}$ with $\delta_{\rm res}(E)$  the FWHM energy resolution taken from Ref.~\cite{2013APh43171B}, while $\Delta E_j$ is the width of the $j$th energy bin. 
 
In the left panel of Fig.~\ref{fig:1}, we show the spectral signature of $\ud \Gamma_{ 1}^{\rm S}/\ud E$ in the first RoI towards the ``classical dSph'' \Dra, for 2 TeV DM particles annihilating, with $\langle \sigma v \rangle = 10^{-24}\,\rm cm^3\,s^{-1}$, into $b\bar{b}$ (black line), $\tau^+\tau^-$ (red line) and $W^+W^-$ (blue line) respectively. From the comparison between the count rate of DM events coming from the GC (central panel in Fig.~2 of~\cite{Lefranc:2015pza}) and those here, we expect that \CTA\ can probe annihilation cross sections of the order of $10^{-23}\,\rm cm^3\,s^{-1}$. Although this value is smaller than the one inferred from a GC study~\cite{Lefranc:2015pza}, the derivation is somehow more robust since the systematic uncertainties on the $J$-factors should be much smaller. Indeed, as pointed out in the previous section, the ON-OFF method is always applicable in dSphs regardless of the DM distribution, while in the GC studies, it can only be performed for DM profiles that feature a significant gradient between the ON and OFF regions (cuspy profiles). The left panel of Fig.~\ref{fig:1} displays the peculiar spectral shape of the predicted DM signals (wide positive features and a sharp peak then cutoff at $E = m_{\rm DM}$).  In particular, a sharp spectral feature occurs for the  DM DM $\rightarrow W^+W^-$ at photon energy close to $m_{\rm DM}$.  This comes from the splitting $W^\pm \rightarrow W^\pm\gamma$ and it becomes more pronounced for very heavy DM candidates.

The right panel of Fig.~\ref{fig:1} shows, instead, the spatial variation of the predicted total $\gamma$-ray rates in \CTA\ summed over all the energy bins above 30 GeV ($\Gamma_{ i}^{\rm S}=1/\Delta t \sum_j N_{ij}^{\rm S}$). These total rates are presented for five dSph galaxies, considering 2 TeV DM particles annihilating into $b \bar b$ with $\langle \sigma v \rangle = 10^{-24}\,\rm cm^3\,s^{-1}$. The predicted count rates for \Dra\ (black-filled dots) and \UMaII\ (cyan circles) are fairly independent of the RoI, since their
$J_i$ show a  shallow modulation with RoI. On  the other hand, the predicted count rates  of \Scu\ (red-filled squares) and \UMi\ (magenta-filled triangles) decrease with RoI, as their $J_i$ values decrease with increasing RoI index, with their very inner count rates expected to be greater than for the first two dSph galaxies.

The right panel of Fig.~\ref{fig:1} clearly indicates that the dSph galaxies in our list should be extended objects for $\gamma$-ray observations with \CTA, which allows measurements of their extent in VHE $\gamma$-rays, if the DM annihilation cross-section is sufficiently large.

\subsection{CTA irreducible background}
\label{Sec:Bkgevents}
As for the other IACTs, the irreducible background (hereafter, background) passing the hardware trigger systems and the analysis cuts is made of cosmic rays (CRs) composed of hadrons (protons, nuclei) and electrons.  Further details on the background and its level in  \CTA\ can be found, for example, in Ref.~\cite{Lefranc:2015pza}. In particular, since the background is isotropic, the number of background photons contained in $\Delta \Omega_i$ is
\beq 
\begin{split}
&N_{ij}^{\rm B}= \Delta t \int_{\Delta E_j} \hspace{-.15cm} \ud E \,
\frac{\ud \Gamma_{i}^{\rm B}}{\ud E} \qquad \hbox {where} \\ 
&\frac{\ud
  \Gamma_{i}^{\rm B}}{\ud E}= \int_{-\infty}^{+\infty} \hspace{-.15cm} \ud E'
\, \frac{\ud \Phi^{\rm CR}}{ \ud E\ud \Omega}(E')
\Delta\Omega^i \, \mathcal A^{\rm CR}_{\rm eff}(E') \,
\frac{\exp\left\{-(E\!-\!E')^2/\left[2\sigma^2(E')\right]\right\}}{\sqrt{2\pi\sigma^2(E')}}
\ , 
\end{split}
\eeq 
and where $\ud \Phi^{\rm CR}/(\ud E\ud \Omega)$ is the total CR background flux per steradian and $\mathcal A^{\rm CR}_{\rm eff}(E)$ is the energy-dependent effective area for CRs. As in our previous study~\cite{Lefranc:2015pza}, the count rate of the  background events $\ud \Gamma_{i}^{\rm B}/\ud E$ is again extracted from Refs.~\cite{2013APh43171B}, taking however into account the smaller angular sizes of  the RoIs considered here with respect to those in Ref.~\cite{Lefranc:2015pza}. This provides the most up-to-date background computation for \CTA\  observations of dSph galaxies.

The left panel of Fig.~\ref{fig:1} shows that the spectral shape of the background in the first RoI (green line) is smoother than the predicted DM signal. We thus foresee a substantial gain in the \CTA\ sensitivity from the potential discrimination power between source and background. This is particularly evident for the $e^+e^-$ and $\mu^+\mu^-$ leptonic channels, as well as the $W^+W^-$ primary mode, since they are characterized by a very prominent feature at $E \simeq m_{\rm DM}$. 

The right panel of Fig.~\ref{fig:1} shows, instead, the spatial variations of the total background  rates, summed over all the energy bins above 30 GeV ($\Gamma_{i}^{\rm B} = 1/\Delta t\sum_j N_{ij}^{\rm B}$).  Contrary to the DM signal in a given dSph, the background increases in proportion to the solid angle size of the RoI. The different spatial trends of the background and the predicted DM signal should allow a gain in sensitivity.

In addition to the background discussed above, one could consider the $\gamma$-ray contamination due to the Galactic Diffuse Emission (GDE) as well. Detailed studies of the GDE expected for \CTA\ have been conducted in Refs.~\cite{Lefranc:2015pza, Silverwood:2014yza}, showing that the GDE contamination could be only important in RoIs close to the GC. Hence, since all dSph galaxies considered in this study are located far away from the Galactic Plane (absolute galactic latitudes greater than $10^\circ$), we expect a negligible  $\gamma$-ray contamination from the GDE.

\subsection{Systematic uncertainties on the $J$-factor}
\label{Sec:SysUncertainties}
Observations of stellar velocities within dSph galaxies allow one to constrain the DM density profiles, through the Jeans equation of local dynamical equilibrium for assumed stationary systems, $ \nabla \cdot {\cal P} = -\nu \,\nabla \Phi $, where ${\cal P} = \nu {\bm \sigma}^2$ is the anisotropic dynamical pressure, $\nu$ is the density of the observed tracer (stars in the dSph), ${\bm \sigma^2}$ is the squared velocity dispersion tensor, while $\Phi$ is the gravitational potential. In spherical symmetry, as nearly always assumed for dSph modeling, the Jeans equation becomes
\begin{equation}
{\ud \left (\nu\sigma_r^2\right) \over \ud r} + 2\,{\beta(r) \over r}\,\left
(\nu\sigma_r^2\right) = -\nu(r) {G\,M(r)\over r^2} \ ,
\label{jeans}
\end{equation}
where $M(r)$ is the total mass (DM, stars, possible gas and black hole),  $\beta = 1 - \sigma_\theta^2/\sigma_r^2$ is  the \emph{velocity anisotropy} (hereafter, anisotropy),  which usually depends on the physical radius $r$ (and where by symmetry, one has $\sigma_\phi = \sigma_\theta$).
The Jeans equation~(\ref{jeans}) contains two unknowns, the mass profile $M(r)$ and the anisotropy profile $\beta(r)$ for a single equation, which is usually called the \emph{mass-anisotropy degeneracy}. There are basically two classes of algorithms to such 
\emph{mass-anisotropy modeling} (see chap. V of  Ref.~\cite{Courteau+14} and references therein):  1) methods based on binning the projected radii and measuring the 0th (surface density), 2nd (lo.s.~velocity dispersion), and possibly 4th (l.o.s.~kurtosis) moment of the l.o.s.~velocity distribution; 2) fitting the distribution of stars in \emph{projected phase space} (projected radii and l.o.s.~velocities).

The computation of the $J$-factors come with numerous systematic uncertainties.
\begin{enumerate}
\item There is no clean method to distinguish member stars from interlopers in the foreground (Milky Way stars), for example using cuts in membership probability or a fully probabilistic membership in a model considering both the dSph stars and Milky Way stars. The inclusion of such interlopers tends to increase the width of the l.o.s. velocity distribution, which in turns leads to greater DM normalizations, hence greater $J$-factors. Furthermore, when they orbit the Milky Way, dSph galaxies are affected by its tidal field, leading to tidal tails, which tend to be elongated towards the center of the Milky Way~\cite{Klimentowski+09_orien},  hence roughly elongated along the l.o.s. given the fairly small distance of the Sun to the center of the Milky Way. This also tends to inflate the l.o.s. stellar  velocity dispersion viewed by an observer on Earth~\cite{Klimentowski+07}. Admittedly,  the alignment of the tidal tail with the l.o.s. is usually not perfect, so the observer should notice an increase of the l.o.s. velocity dispersion at large projected radii. Different methods of handling interlopers (removing them or including them in a probabilistic fashion) lead to typical differences in the $J$-factor of 0.1 dex,  especially in poorly-sampled dSph galaxies at low galactic latitudes (i.e. with important contamination by Milky Way foreground stars)~\cite{Bonnivard+16}.

\item There are several stellar populations in many dSph galaxies, each with different distributions in projected phase space~\cite{WP11,AE12}, each with its own Jeans equation~(\ref{jeans}) probing the same gravitational potential. Neglecting these differences by considering a single stellar population will decrease the accuracy of the derived $J$-factors, perhaps by 0.3 dex.

\item Many authors bin their data, but the mass-anisotropy modeling can be affected to the point of concluding to either cusp or core for the inner DM density profile depending on the choice of binning scheme \citep{RF14}. This should lead to differences of at least 1 dex in the $J$-factors.

\item All mass-anisotropy modeling studies of dSph galaxies assume spherical symmetry, although these galaxies  appear to be slightly flattened along the line-of-sight. The assumption of spherical symmetry leads to typical overestimates of $J$ by 0.5 dex \citep{BCMW15} and probable uncertainties of 0.2 dex.

\item Nearly all mass-anisotropy modeling assumes constant velocity anisotropy ($\beta$ in eq.~[\ref{jeans}]), while one expects that the outer regions should have more radial orbits, as inferred in giant elliptical galaxies from  simulations~\cite{Dekel+05}. For mock dSph galaxies with isotropic orbits inside and strongly radial orbits outside,  the assumption of constant velocity anisotropy leads to overestimates of $J$ as high as 0.7 dex \citep{BCMW15} and probable uncertainties of 0.2 dex.

\item The stellar number density profiles ($\nu(r)$ in eq.~[\ref{jeans}])  of dSph galaxies are not well known because their overall surface densities are low and difficult to distinguish against the foreground Milky Stars, even after selecting with colors and metallicities. This should produce uncertainties of 0.2 dex in the $J$-factors.

\item Even the center of a classical dSph like Fornax is uncertain by a few arc minutes \citep{Battaglia+06}. An incorrect center should lead to underestimates on the cuspiness of both stellar and DM density profiles, hence on underestimated $J$-factors. The  effect of incorrect centering on the $J$-factor still needs to be quantified, and we estimate its level at 0.2 dex uncertainty in the $J$-factor.

\item The usual mass-anisotropy modeling of dSph kinematics in the context of $J$-factors  incorrectly assumes that the l.o.s. velocity distribution is Gaussian (Maxwellian), when we know from mass/orbit modeling theory that velocity anisotropy creates non-Gaussian distributions for $v_{\rm los}$ \citep{Merritt87}, so that the shape of the l.o.s. velocity distribution actually helps to lift the mass-anisotropy degeneracy of the Jeans equation~(\ref{jeans}) and thus obtain more accurate masses. We estimate that this induces another 0.1 dex uncertainty in the $J$-factor.

\item When the data are binned to obtain l.o.s. velocity dispersion profiles, it is formally incorrect to perform a $\chi^2$ analysis, since the distribution of velocity dispersions is not Gaussian, even if the l.o.s. velocity distribution were Gaussian (see previous item). Instead, the distribution of velocity dispersions has wide tails, since the distribution of squared velocity dispersions follows the wide-tailed $\chi^2$ distribution. Again, we estimate that this induces a 0.1 dex uncertainty in the $J$-factor.

\item Very few authors consider the effects of tidal stripping on the DM halos of dSph galaxies. Although this mainly affects the decay factor (which is the analogous to the $J$-factor, but with a power of unity of the density of DM in eq.~[\ref{Jdef}]) instead of a power two), it also has a small but non-negligible effect on the  $J$-factor, especially at large projected radii, which we estimate as a 0.3 dex overestimate.

\item Few authors consider the central supermassive black holes (BHs) that many expect to reside in the cores of dSph galaxies, if these are the small-scale analogs of giant galaxies that are known to harbor massive singularities in their centers. BHs have two effects on the $J$-factors. First, they attract DM, making it much more cuspy (with a density slope steeper than $-2$, \citep{QHS95}), which leads to a boost of typically 1 dex in $J_{\rm ap}$, and up to 6 dex in some cases  \citep{GPQ14}. On the other hand, the presence of a BH will lead to reduced overall DM in the mass-anisotropy modeling, thus reducing the $J$-factor.  The effects of the BH are all the more important that in tidally stripped systems such as dSph galaxies, the ratio of BH mass to stellar mass will be perhaps 10 times greater than in galaxies of the same mass that lie well beyond the virial radius of their nearest major galaxies.
\end{enumerate}

One can attempt to estimate the global systematic uncertainty in $J_{\rm ap}$ at a radius of two half-light radii, where many of the individual contributions to the systematic uncertainty have been derived in Ref.~\cite{BCMW15}.  The uncertainties of the 11 causes of systematic errors add up in quadrature to 0.57 dex. Moreover, the biases may not be easy to correct for, in which case, adopting bias/$\sqrt{12}$ as the uncertainty for each and summing up in quadrature, we have an additional uncertainty of 0.49 dex, for a total systematic uncertainty of 0.76 dex.

These systematic biases and uncertainties are expected to be worse for the $J_i$  than for $J_{\rm ap}$, since the latter is an integral of the former so that some radially dependent systematics will be washed out. Hence, we estimate that the systematic uncertainties on the $J_i$ should be $\gtrsim 1$ dex.

\section{Analysis methodology}
\label{Sec:AnaMet}
To derive the \CTA\ sensitivity, we perform a likelihood ratio statistical test, which allows us  to take full advantage  of the spectral and spatial morphologies of the DM signal compared to the background as shown in Fig.~\ref{fig:1}. For a given $m_{\rm DM}$, the total likelihood for the  $k$th dSph galaxy is obtained as  the product of  two likelihood terms over the spatial bins $i$ and energy bins $j$. It is given by 
\begin{equation}
\label{totlik}
\mathcal{L}_k  (\mDM,\langle \sigma v \rangle) = \prod_{i,j} \mathcal{L}_{k\, ij}  (\mDM,\langle \sigma v \rangle)  \mathcal{L}_{k\, i}^J   \ ,
\end{equation}
where the energy bins are logarithmically spaced between 30 GeV and 80 TeV and the spatial ones have a width of 0.1$^{\circ}$. The procedure for deriving the number of RoIs for each dSph considered in this study is explained in  Sec.~\ref{Sec:RoIs}. The  likelihood term $\mathcal{L}_{k\, ij}  (\mDM,\langle \sigma v \rangle)$ explicitly writes~\cite{2011EPJC711554C}
\begin{equation}\label{Likelihood}
{\mathcal{L}_{k \, ij}(N_{}^{\rm S},N_{}^{\rm B}|N_{\rm ON},N_{\rm OFF}) = \frac{(N_{ k \, ij}^{\rm S}+N_{ ij}^{\rm B})^{N_{{\rm ON}, k \, ij}}}{N_{{\rm ON},k \, ij}!}e^{-(N_{k \, ij}^{\rm S}+ N_{ ij}^{\rm B})} \frac{\left(N_{ ij}^{\rm B}/ \alpha_i \right)^{N_{{\rm OFF},ij}}}{N_{{\rm OFF},ij}!} e^{-N_{ ij}^{\rm B}/\alpha_i}} \ ,
\end{equation}
where the parameter $\alpha_i=\Delta \Omega_\odot^i/\Delta \Omega_{\rm OFF}$ is the ratio between the $i$th ON and OFF region angular sizes.  Here, $N_{{\rm ON},k \, ij}$ and $N_{{\rm OFF}, ij}$ are the observed number of events in the ON and OFF regions respectively which correspond to Poisson realizations with mean  $N_{ k\, ij}^{\rm S} + N_{ ij}^{\rm B}$ and $N_{ij}^{\rm B}/\alpha_i$ (see Ref.~\cite{Lefranc:2015pza} for further details). The background is estimated in the OFF region,  which corresponds
to an annulus covering part of the \CTA\ FoV (around  7.5$^\circ$ for the middle size telescopes). For all  dSph galaxies considered here, the OFF region is at least five times bigger than the last ON region, as one can see either in the right panel of Fig.~\ref{fig:1} or more specifically in the fifth column of Tab.~\ref{tab:dSphs}. Hence $\alpha_i \leq 0.2$.  

In Eq.~\eqref{totlik}, the likelihood term $\mathcal{L}_{k\, i}^J$ measures the impact of the statistical uncertainties on the $J$-factor derivations.  For the $k$th dSph galaxy and $i$th RoI, it explicitly writes~\cite{Ackermann:2011wa} 
\begin{equation}\label{JLikelihood}
\mathcal{L}_{k\, i}^J  (J | \tilde J_i^k,\sigma_i^k) = \frac1{\ln 10  \,
  \tilde J_i^k} \, \mathcal G (\log J | \log \tilde J_i^k,\sigma_i^k) \ ,
\end{equation}
where  $\tilde J_i^{k}$ is the observed $J$-factor of galaxy $k$ in the $i$th annulus, $\sigma_i^k$ is the uncertainty on $\log \tilde J_i^k$ (both determined from kinematical modeling by Ref.~\cite{Geringer-Sameth:2014yza}), and $\mathcal G(x|\mu,\sigma)$ is the Gaussian distribution of mean $\mu$ and standard deviation $\sigma$. The procedure for deriving $\tilde J_i^{k}$ and $\sigma_i^k$  in our RoIs is explained in Sec.~\ref{Sec:AnalysisSetup}.  

\smallskip
In what follows, in order to estimate which is the impact of both the spectral and morphological analysis in the \CTA\ sensitivity, we will adopt three statistical approaches:
\begin{itemize}
\item[$\diamond$] {\it integrated approach:} The number of predicted signal and irreducible background events  are given by summing over all energy and spatial bins. As a consequence, one has to  modify Eq.~\eqref{Likelihood} by replacing $N_{k \, ij}^{\rm S} \rightarrow N_{ k}^{\rm S} \equiv \sum_{i,j} N_{ k \, ij}^{\rm S}$ and $N_{  ij}^{\rm B} \rightarrow N^{\rm B} \equiv \sum_{i,j} N_{  ij}^{\rm B}$ and Eq.~\eqref{JLikelihood} by substituting $ J^k_i \rightarrow  J^{k} $ and $\sigma_i^{k} \rightarrow \sigma^{k}  $. The values of $ J^{k  } $ for our set of dSphs are given in the last column of Tab.~\ref{tab:dSphs}. For each dSph, there is then just one ON region that covers part of the angular size of the object,  and one OFF region, with $\alpha=0.2$. This is clearly the most conservative approach. Indeed, neither the peculiar spectral features  of the DM signal with respect to the irreducible background nor the morphological ones can be used to improve the \CTA\ sensitivity. 

\item[$\diamond$] {\it spectral only approach:} We sum over all spatial bins. As a result, the individual likelihoods in Eq.~\eqref{Likelihood} are obtained by replacing $N_{ k \, ij}^{\rm S} \rightarrow N_{ k \, j}^{\rm S} \equiv \sum_{i} N_{ k \, ij}^{\rm S}$ and $N_{ ij}^{\rm B} \rightarrow N_{ j}^{\rm B} \equiv \sum_{i} N_{  ij}^{\rm B}$. The likelihood term in Eq.~\eqref{JLikelihood} is determined as described in the previous item. Unlike the {\it integrated approach}, one can now profit of the bump-like features and sharp cut-off of the DM signal in order to ameliorate the \CTA\ sensitivity. 

\item[$\diamond$] {\it 2D approach:} We directly use  the  likelihood terms for each energy and spatial bin defined in Eqs.~(\ref{Likelihood}, \ref{JLikelihood}). This approach, in addition to the spectral features, lets us also to take into account for the different spatial behaviour of the DM signal compared to the background. 
\end{itemize} 

\smallskip
Furthermore, it would be also relevant to assess whether a joint analysis of observations towards a selected set of dSphs can be used to improve the \CTA\ sensitivity. This might be important for two main reasons: $i)$  the \CTA\ observatory will probably dedicate significant observation time to the most promising dSph galaxies at the first three years of operation; $ii)$ the forthcoming extragalactic survey will provide  an unbiased picture of a quarter of the sky at a sensitivity of several mCrab level~\cite{2013APh43317D}.  Such a survey will be particularly interesting in DM context, since a collection of dSphs galaxies will be present in the observed \CTA\ FoV.  Therefore, when a collection of dSphs is considered, our statistical approach will be based on a  joint binned Poisson likelihood. It writes 
\begin{equation}
\mathcal{L} (\mDM,\langle \sigma v \rangle) = \prod_{k} \mathcal{L}_{k}  (\mDM,\langle \sigma v \rangle) \ .
\end{equation}

\smallskip
The statistical analysis proceeds with a likelihood ratio Test Statistic (TS). Schematically, for each of the statistical methods described above,   $\rm TS =  -2\ ln[\mathcal{L}(m_{\rm DM},\langle \sigma v \rangle) / \mathcal{L}_{\rm max}(m_{\rm DM}, \langle \sigma v \rangle )]$. This, for a given $m_{\rm DM}$,  follows an approximate $\chi^2$ distribution with one degree of freedom ($\langle \sigma v \rangle$). Values of TS higher than 2.71 are excluded at a 95\% Confidence Level (C.L.). Further details on the development of statistical methods in VHE astrophysical searches for particle DM can be found, for instance, in Ref.~\cite{Conrad:2014nna}.

\section{Results}
\label{Sec:Results}

\begin{figure}[!t]
\centering
\includegraphics[width=.495\textwidth]{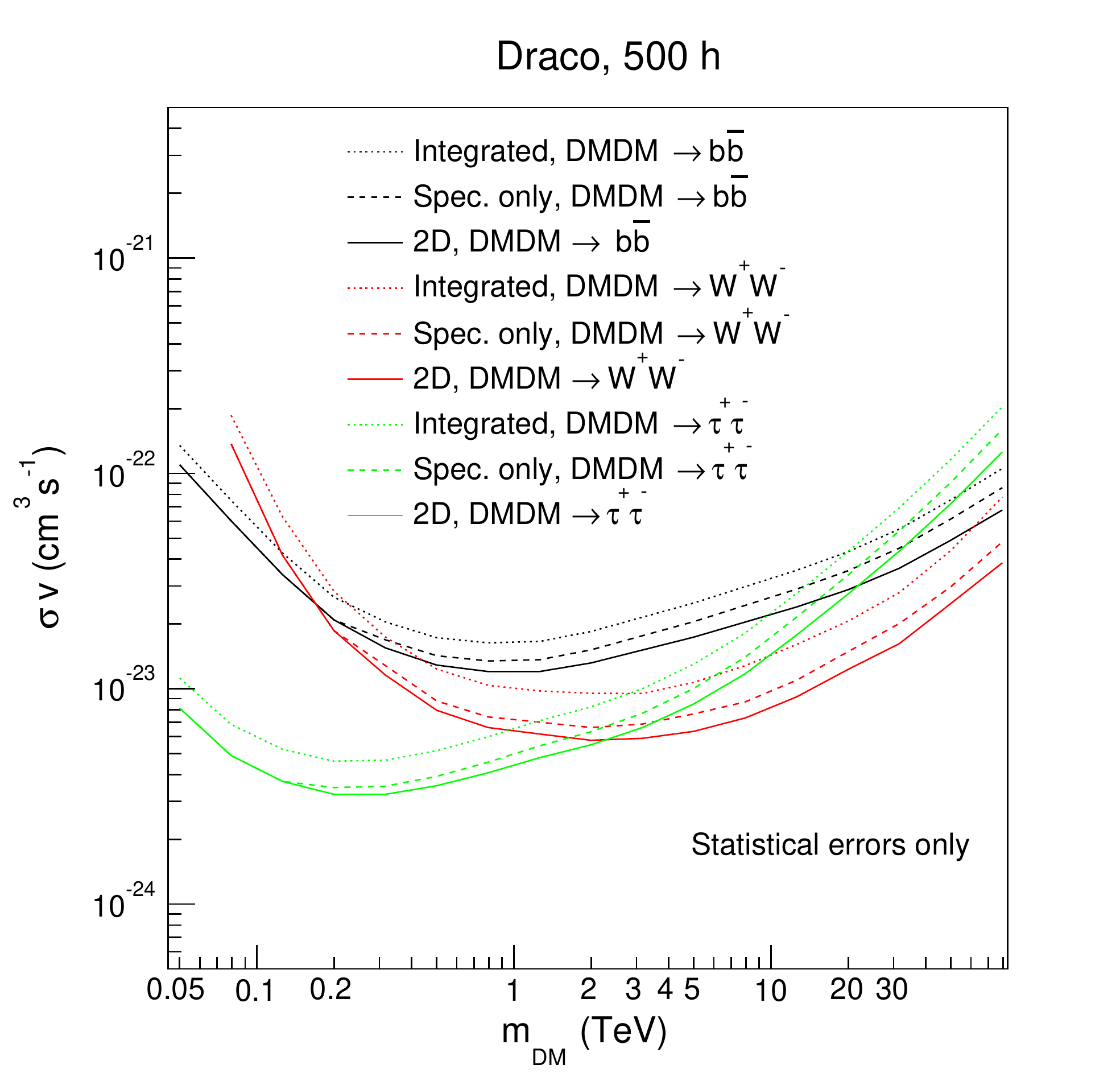}
\includegraphics[width=.495\textwidth]{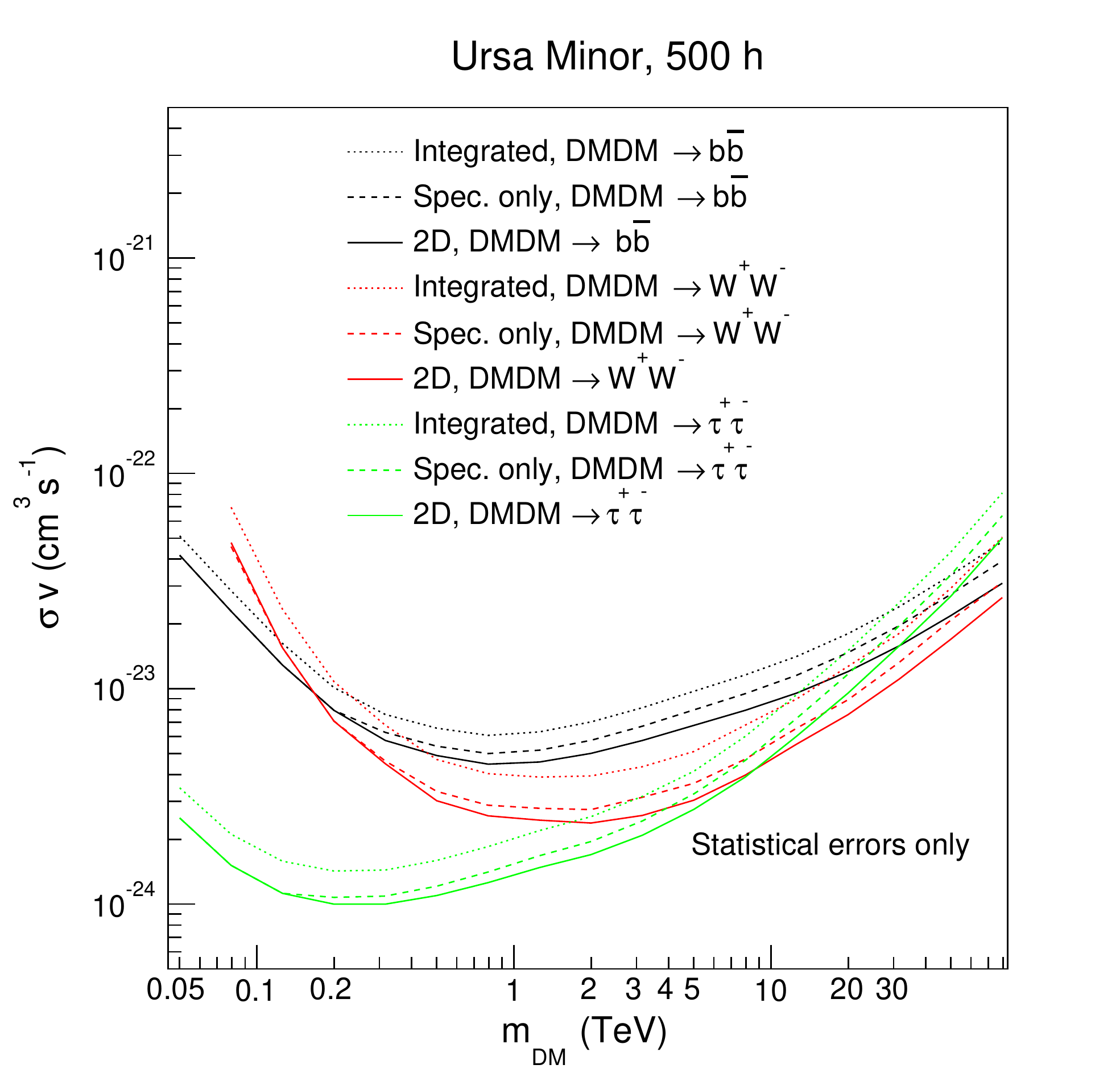}
\includegraphics[width=.495\textwidth]{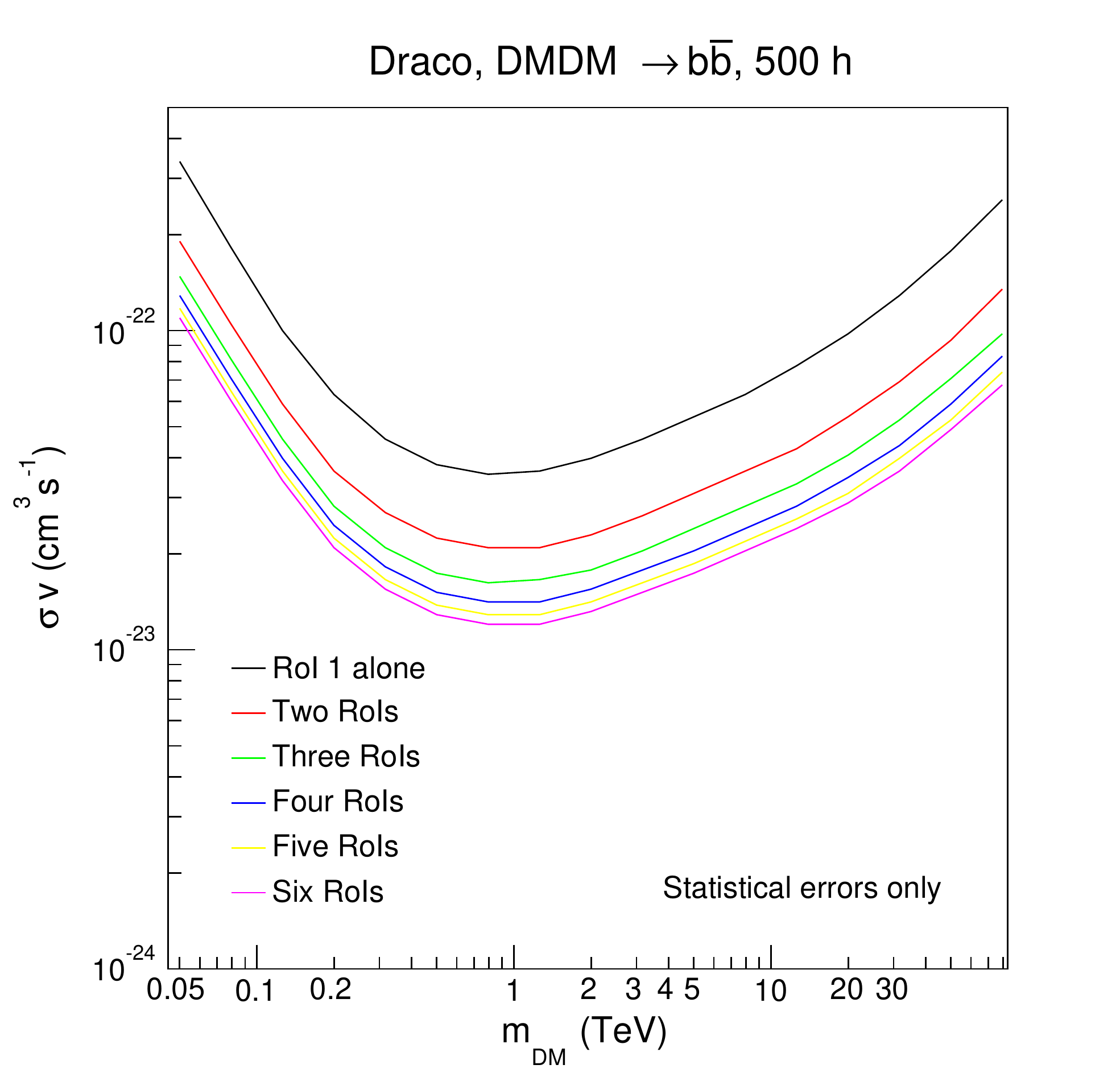}
\includegraphics[width=.495\textwidth]{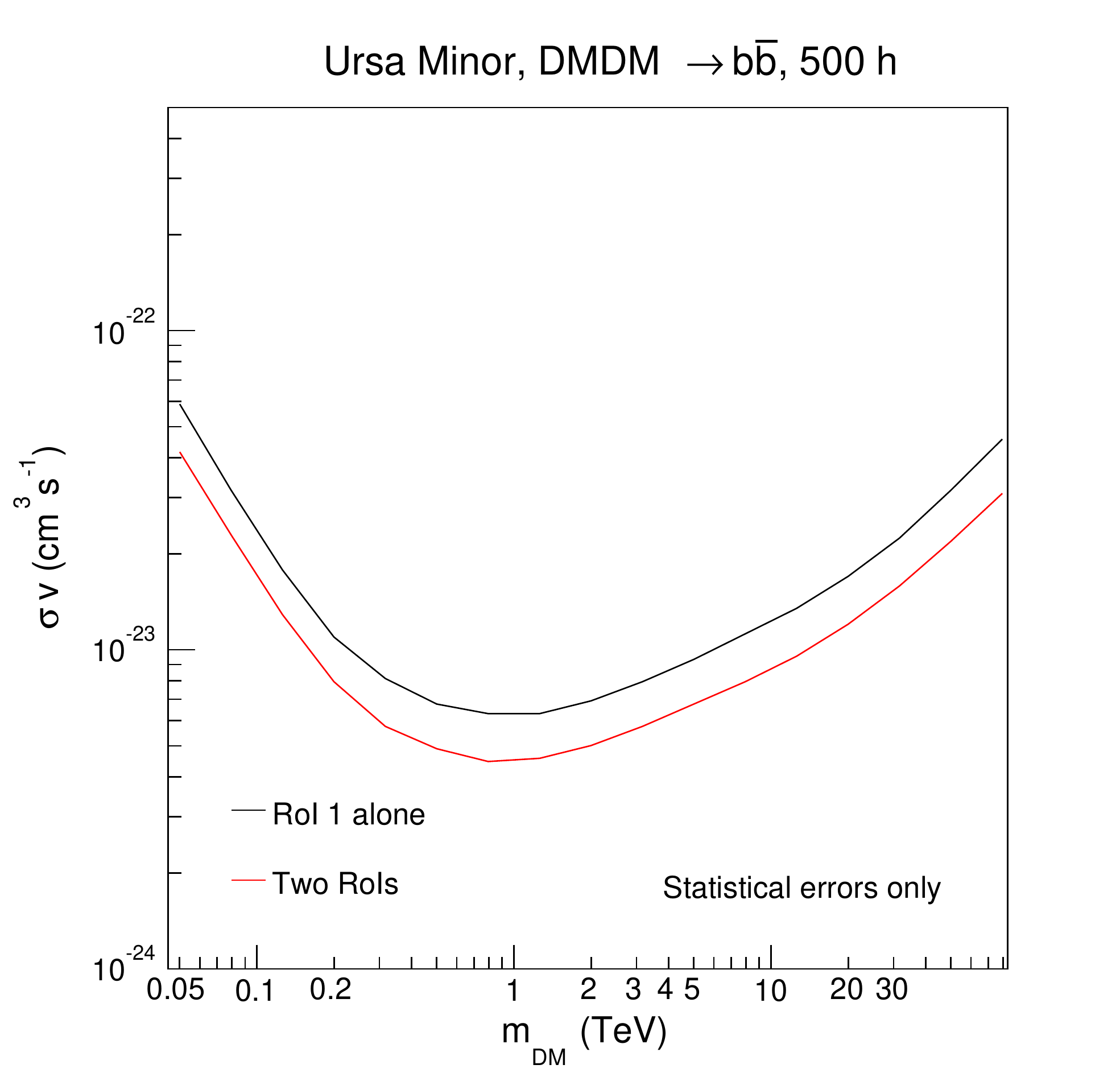}
\caption{\small CTA sensitivity to DM annihilation in the ($\mDM, \sigmav)$ plane. The analysis employs 500 h exposure in each RoI, an energy threshold of 30 GeV, and we only consider  the statistical uncertainties as described  in Sec.~\ref{Sec:AnaMet}. {\it Upper panels}: Impact of the different  statistical approaches  on the \CTA\ sensitivity towards \Dra\ (left panel)  and \UMi\ (right panel) for the DM DM $\rightarrow b \bar{b}$ (black  lines), $W^+W^-$ (red lines) and $\tau^+\tau^-$ (green lines) primary channels. In particular, the solid lines are obtained by implementing a  full 2D statistical approach, while the dashed and dotted ones respectively only use the spectral information (no gain in sensitivity from the spatial morphologies of the DM signal) and integrated (no gain in sensitivity from both the spatial and spectral morphology of the DM signal) approaches. {\it Lower panels}: Improvement of the \CTA\ sensitivity towards \Dra\ (left panel) and \UMi\ (right panel) adding all the available RoIs  for a signal-to-noise ratio optimization. See Sec.~\ref{Sec:AnaMet} for further details. }
\label{fig:2}
\end{figure}

\begin{figure}[t]
\centering
\includegraphics[width=.3275\textwidth]{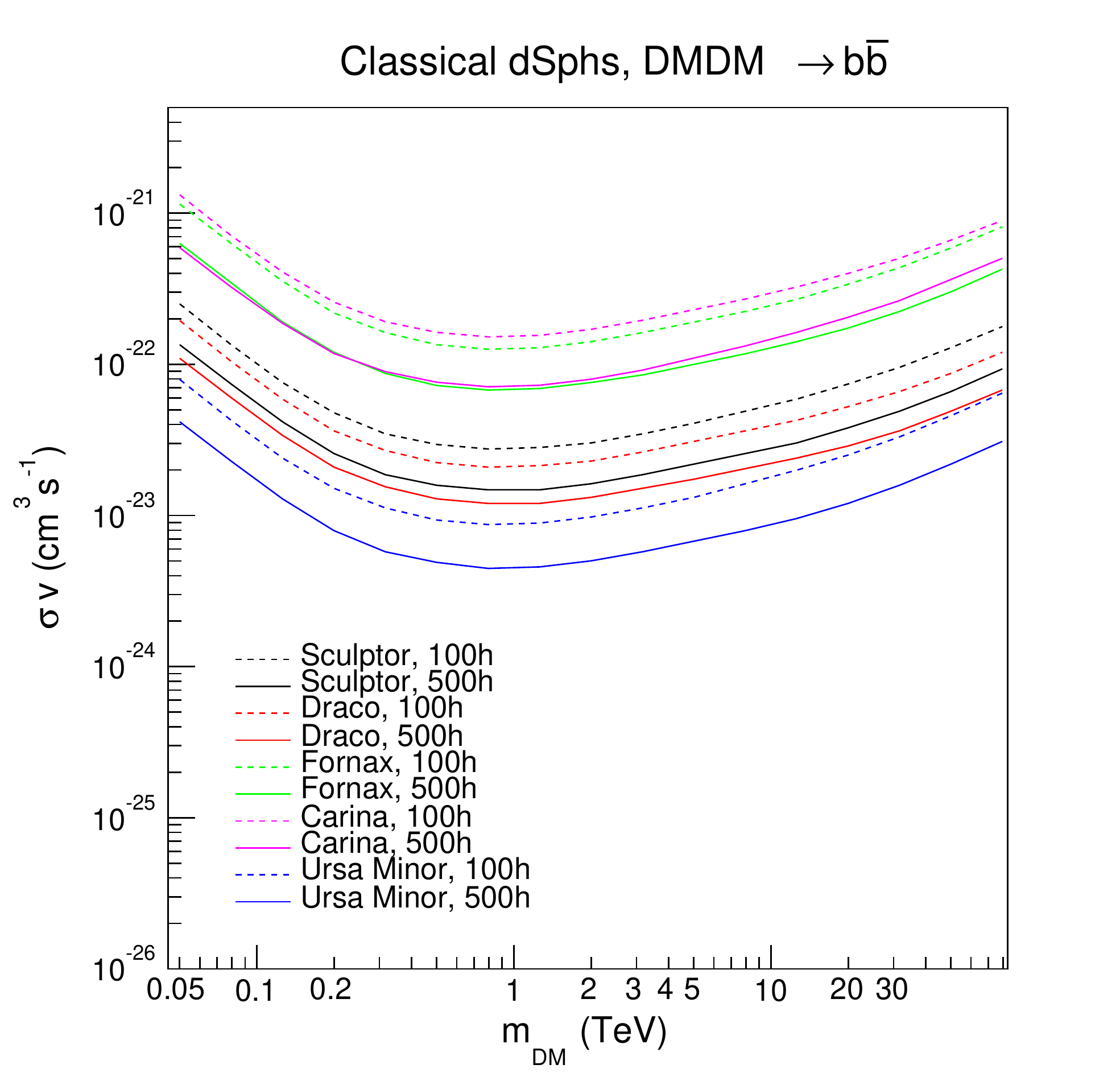}
\includegraphics[width=.3275\textwidth]{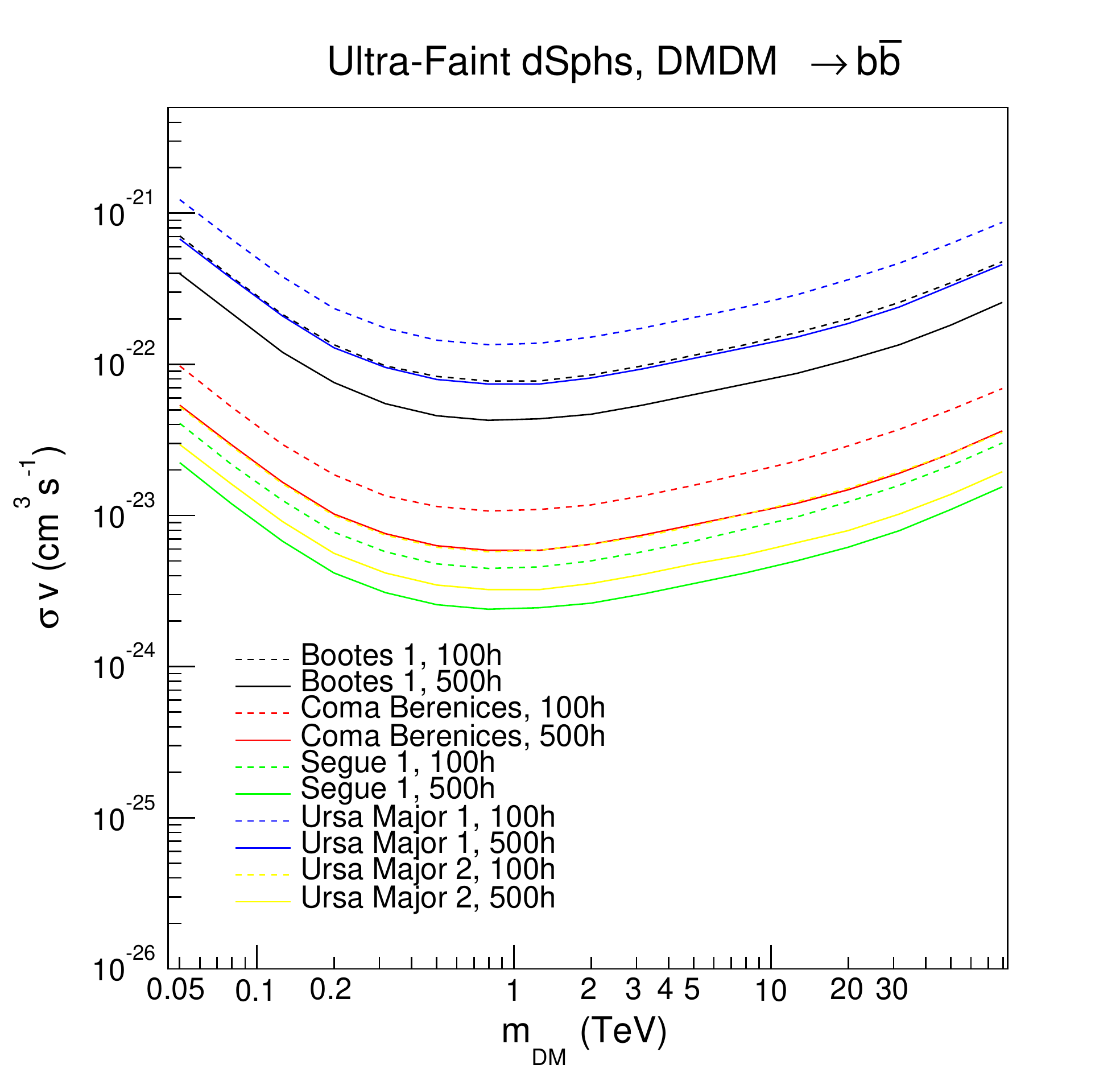}
\includegraphics[width=.3275\textwidth]{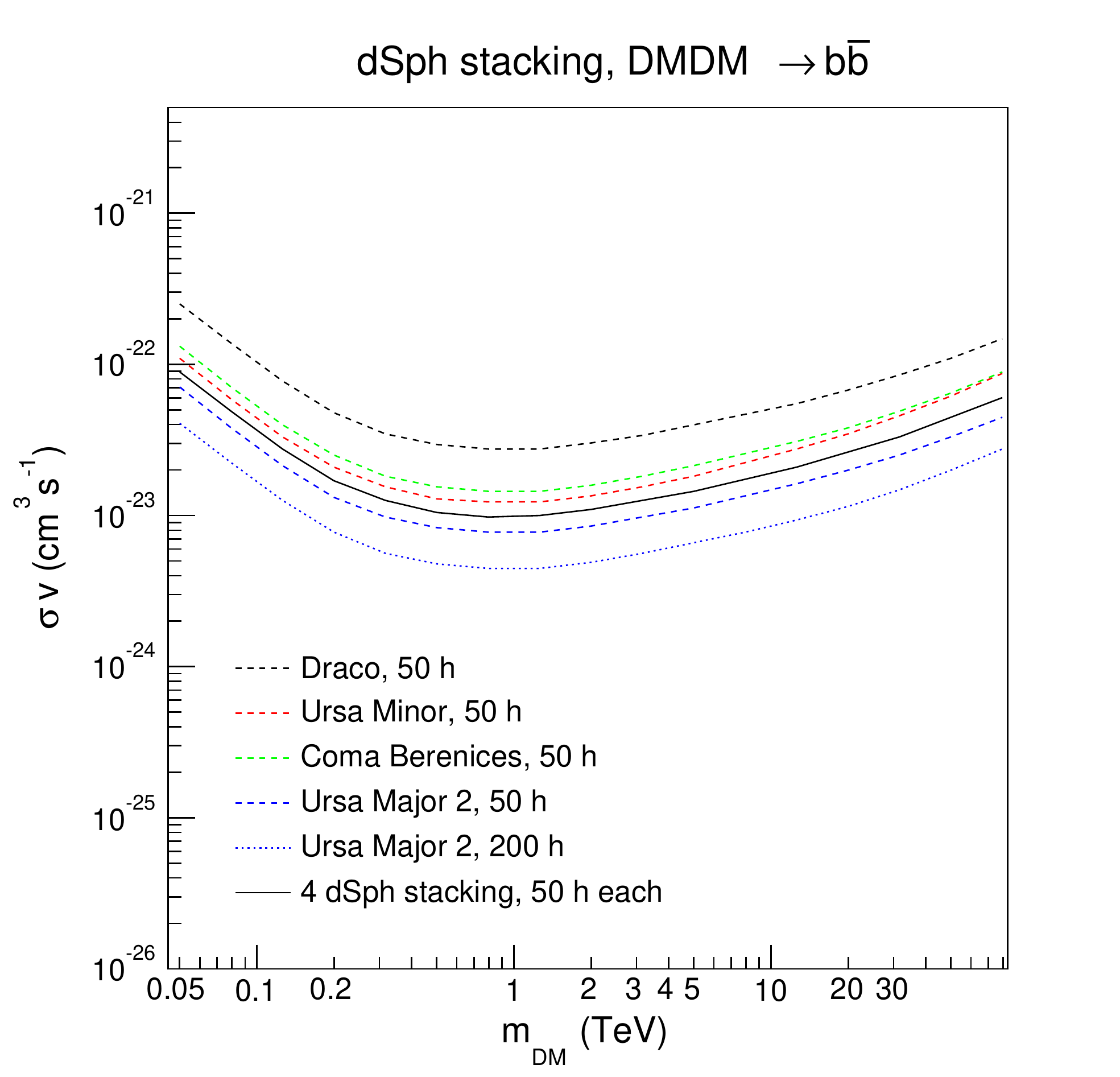}

\includegraphics[width=.3275\textwidth]{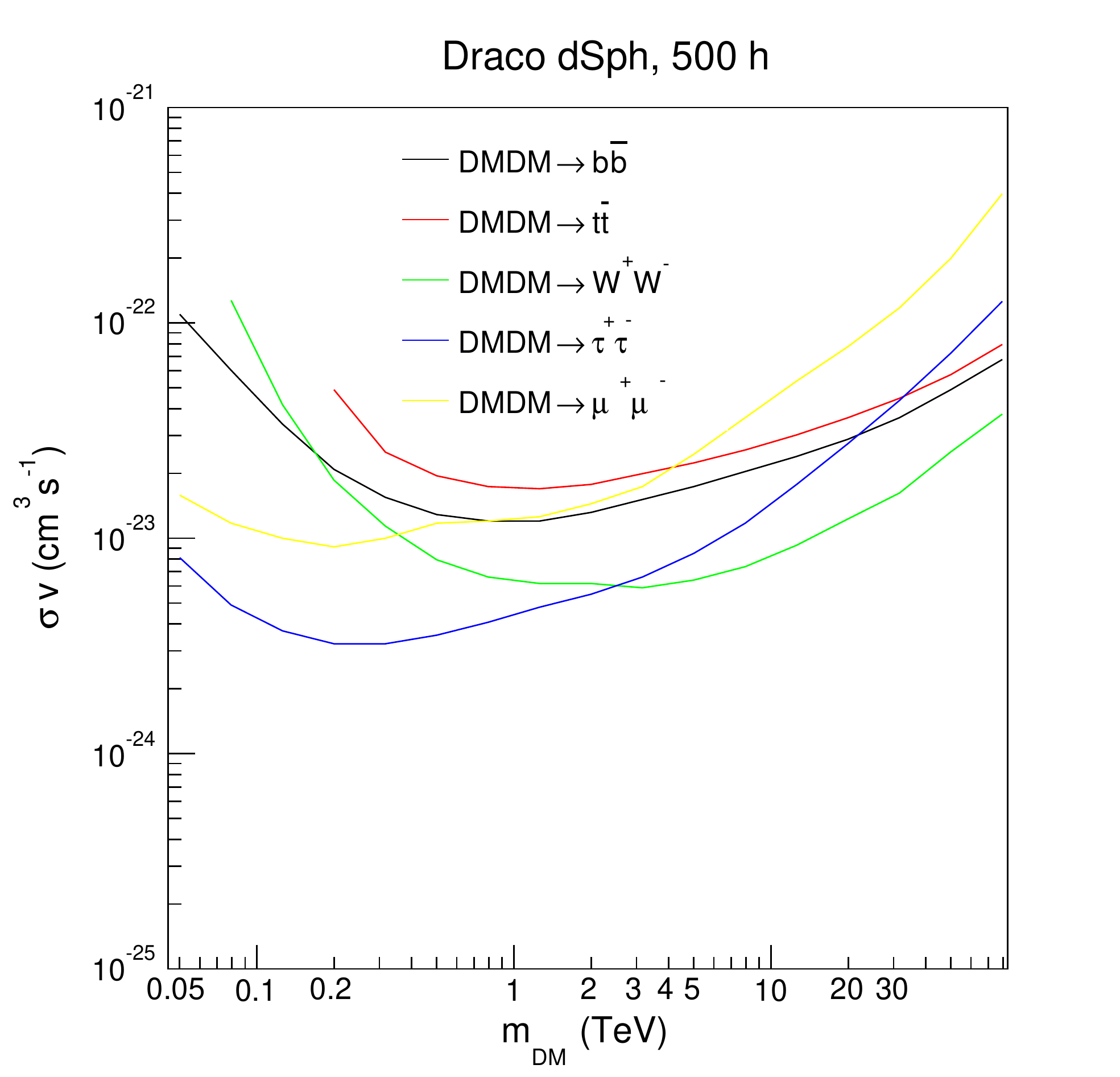}
\includegraphics[width=.3275\textwidth]{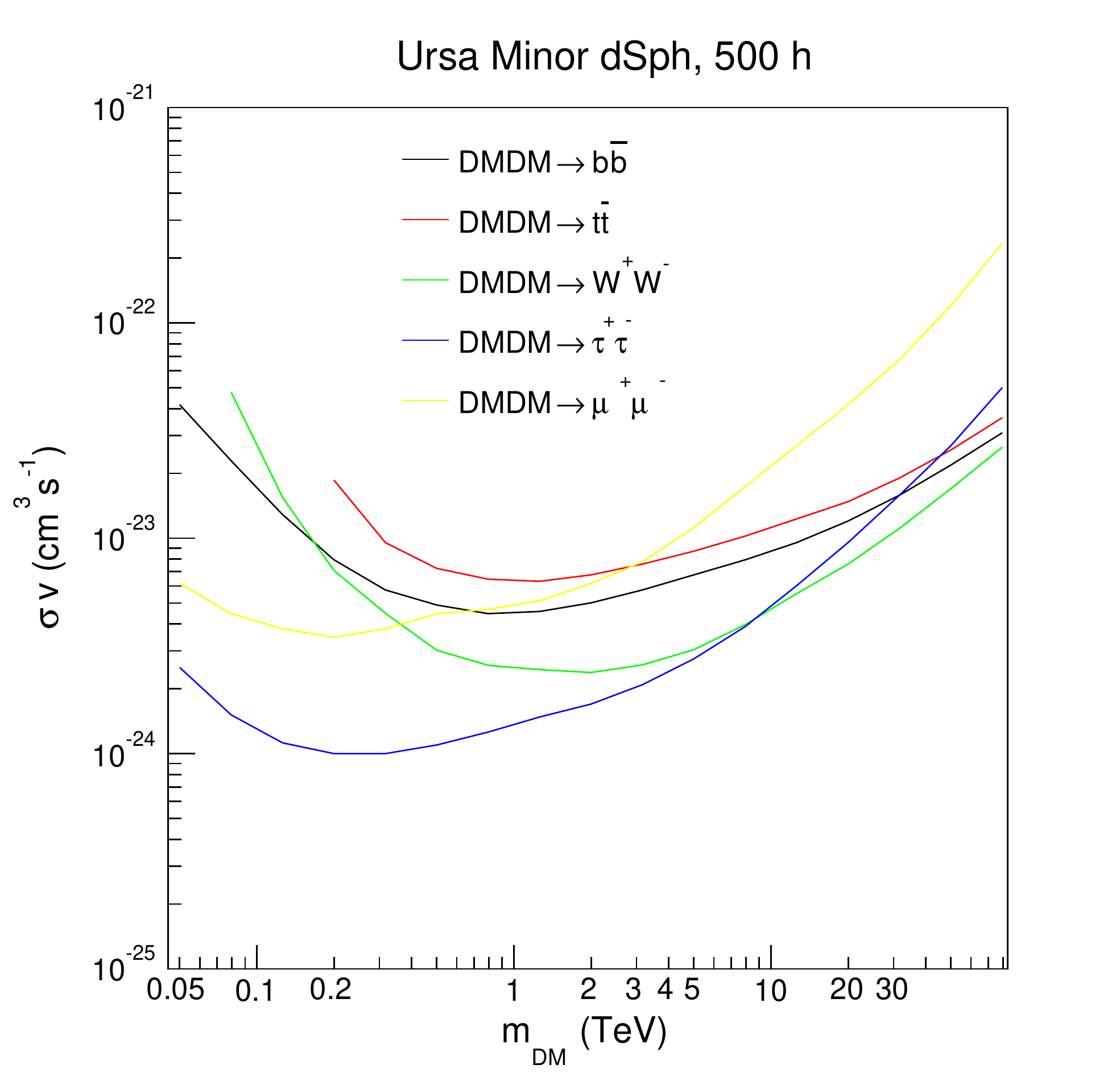}
\includegraphics[width=.3275\textwidth]{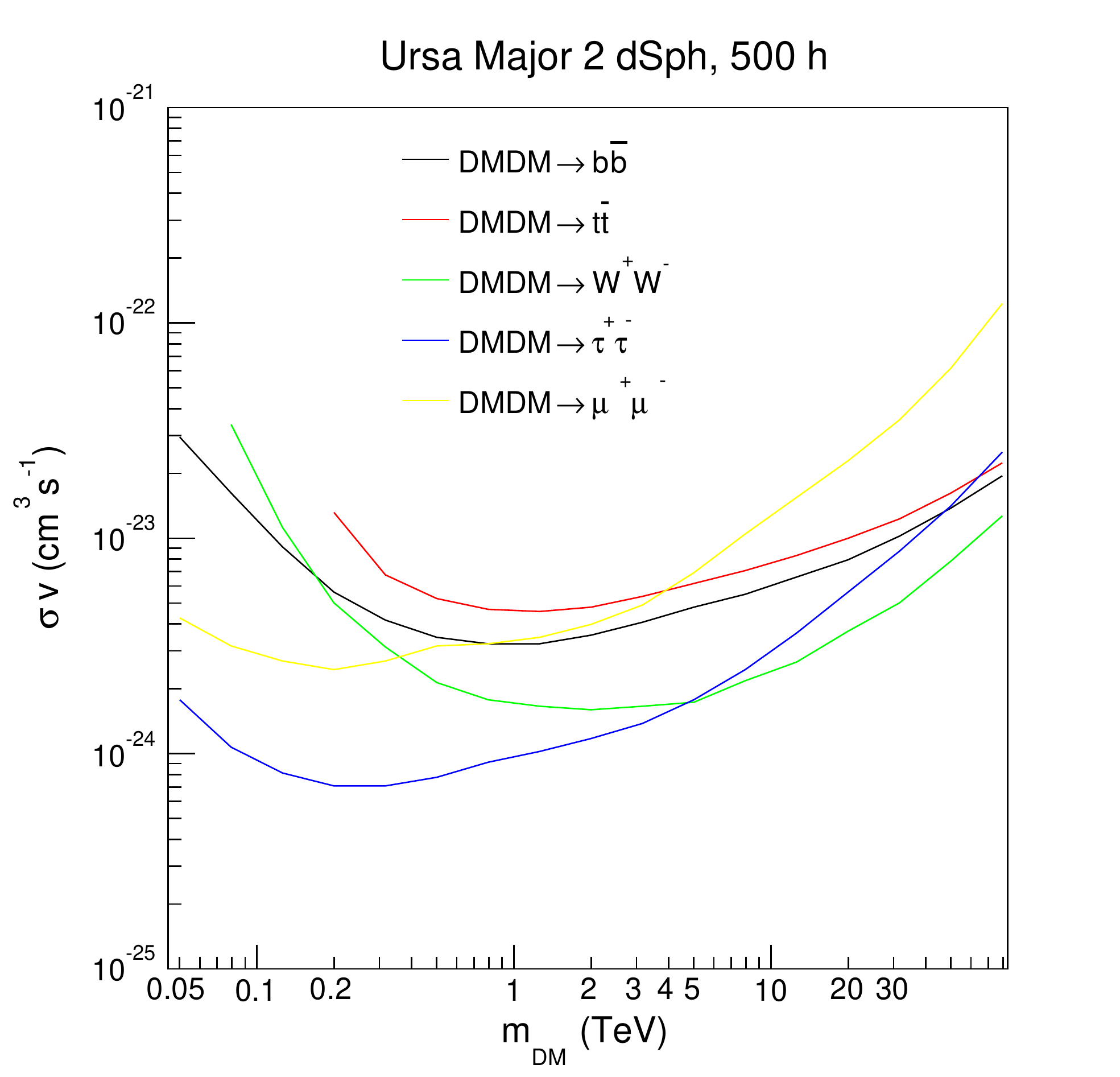}
\caption{\small  \CTA\ sensitivity obtained by implementing a 2D statistical analysis. {\em Upper panels:} In the left and central panels we show the sensitivity for our list of ``classical'' dSphs and ``ultra-faint'' ones for 100 h (dashed lines) and 500 h (solid lines) of observations respectively. In the right panel, we report the bounds from a combined  analysis of time-limited observations of 4 dSph galaxies (\Dra, \UMi, \Com\ and \UMaII) as well. In all panels, the sensitivity is shown for  the DM DM $\rightarrow b \bar b$ primary annihilation channel. {\em Lower panels:}  Sensitivity to DM annihilation towards \Dra\ (left panel), the best ``classical'' dSph \UMi\ (central panel) and the best ``ultra-faint'' one \UMaII\  (right panel).  We consider several primary channels: DM DM  $\rightarrow  b\bar b$ (black lines), $t \bar t$ (red lines),  $W^+W^-$ (green lines), $\mu^+\mu^-$ (yellow lines), and $\tau^+\tau^-$ (blue lines),  and an observation time of 500 h. See the text for further details. }
\label{fig:3}
\end{figure}

We now present  the $95\% \rm{\ C.L.}$ \CTA\ sensitivity to DM annihilation in the ($m_{\rm DM}$,$\sigmav$) plane. We focus on a broad range of DM masses (from 50 GeV up to 80 TeV) and several annihilation primary modes (DM DM $\rightarrow$ $\mu^+\mu^-$, $\tau^+\tau^-$, $b\bar{b}$, $t\bar{t}$, and $W^+W^-$). For all  channels, we improve the \CTA\ sensitivity to DM annihilation by taking advantage of the peculiar features of the DM signal with respect to the irreducible background.   We consider a large sample of dSph galaxies that have the most promising $J$-factors and smaller statistical uncertainties accordingly with Refs.~\cite{Geringer-Sameth:2014yza, Bonnivard:2015xpq, Bonnivard:2015tta}. For all  considered dSph galaxies, we implement   a multi-bin morphological method that further improves the \CTA\ sensitivity. We summarize in Fig.~\ref{fig:2} our main results. We find that:
\begin{itemize}
\item {\it A spectral statistical approach substantially increases the CTA  sensitivity:} Assuming a 500 h observation time, in the upper panels of  Fig.~\ref{fig:2}, we show the improvement of the \CTA\ sensitivity in  different annihilation channels by using a spectral statistical approach  (dashed lines) compared to a case where only an integrated method is  implemented (dotted lines). The dSph galaxies considered are  \Dra\ (upper  left) and \UMi\ (upper right). For both  integrated and spectral  statistical approaches, we  use  one ON region with ${\rm log}_{10}  J^k =  18.89^{+0.14}_{-0.14}$ for \Dra\ and  ${\rm log}_{10}  J^k =  18.89^{+0.29}_{-0.29}$ for \UMi. As one can see,  we get the maximal  improvement ($\simeq 40\%$) for the DM DM $\rightarrow W^+W^-$ channel (red  lines), since the DM signal is characterized by a prominent spectral  feature close to $m_{\rm DM}$ (see the left panel of Fig.~\ref{fig:1}). For  the DM DM $\rightarrow \tau^+\tau^-$ mode (green lines) the sensitivity  increases by $\simeq30\%$, while  for the DM DM $\rightarrow  b \bar b$  primary channel we get a  $\simeq20\%$ improvement, since the DM signal is then very  broad in energy without any peculiar spectral feature (see again the left panel of Fig.~\ref{fig:1}). 

\item {\it A 2D statistical method further improves the CTA sensitivity:} The upper panels of Fig.~\ref{fig:2} also show the improvement of our 2D statistical method (solid lines). More specifically, in addition to the above mentioned spectral approach in which only one ON region covering the angular sizes of \Dra\ and \UMi\ is used, we implement a multi-bin morphological  analysis that takes advantage of the spatial extent of such dSphs. As is apparent, the \CTA\ sensitivity further improves of an additional $\simeq10\%$  at high DM masses, regardless of the channels, while there is no improvement for $m_{\rm DM} \lesssim 200$ GeV.  More specifically, the lower panel of the same figure shows, again for \Dra\ (lower left) and \UMi\ (lower right), how the \CTA\ sensitivity in the DM DM $\rightarrow b \bar b$ mode increases adding all the available RoIs for a signal-to-noise ratio optimization. As one can see, one can use more RoIs for \Dra, as the DM spatial profiles are more spread. Finally, it worth stressing that  the \CTA\ sensitivity towards the two dSphs is not the same even though the DM signals are enhanced by the same $J$-factor (see Tab.~\ref{tab:dSphs}). This is due to the fact that the $J$-factor profile of \Dra\ is extended in a region twice the angular extent as that of \UMi. As a consequence, the
spatial integration for \Dra\ leads  to a higher residual background that degrades the \CTA\ sensitivity. 

\item {\it The dSph galaxies are extended objects for CTA:} This can be seen  either  in the lower panel of Fig.~\ref{fig:2} or directly in  Tab.~\ref{tab:dSphs}. In particular, the dSph galaxies with flat $J$-factor  profiles (e.g.~\Dra, \UMaII) are more extended objects in $\gamma$-rays  than dSphs with steep profiles (e.g.~\UMi\ and \Com). Furthermore, it is worth noticing  that the dSph galaxies considered in this study have a  smaller spatial extent in $\gamma$-rays than the angular sizes of the objects. This is due to the fact that the irreducible background in the outer RoIs is so large that it completely overshadows the feeble DM signal. 
\end{itemize}

In the upper-left and -central panels of Fig.~\ref{fig:3}, we show the bounds   in the ($m_{\rm DM}$,$\sigmav$) plane for the DM DM $\rightarrow b \bar b$ primary channel for our list of ``classical'' dSphs (left panel) and ``ultra-faint'' ones (central panel).  The projections are obtained by pointing the \CTA\ array for 100 h (dashed lines) and 500 h (solid lines) of observation towards  a given dSph. 

\smallskip
One can see that for the  ``classical'' dSphs, the highest sensitivity is achieved by observing \UMi\ (blue lines). In particular,  the sensitivity is $\langle \sigma v \rangle \lesssim 4.7 \times 10^{-24}$ cm$^3s^{-1}$ at $m_{\rm DM}\simeq1$ TeV and 500 h of obervation. For the other dSph galaxies the bounds are always weaker than  $\langle \sigma v \rangle \lesssim 10^{-23}$ cm$^3$s$^{-1}$. In summary, given its high DM content, relatively peaked $J$-profile and small statistical uncertainties, \UMi\ is the best ``classical'' dSph  for the study of VHE $\gamma$-rays from  annihilating DM by  \CTA.  

\smallskip
In the upper-central panel of the same figure, we also report the sensitivity for ``ultra-faint'' dSph galaxies.  In some cases, the mean $J$-factors are $\mathcal O(10)$ higher than the one of \UMi\ (see the last column in Tab.~\ref{tab:dSphs}). For example, for \Seg\ and \UMaII, the total $J$-factors summed over the relevant RoIs are ${\rm log}_{10}  J^k = 19.33^{+0.32}_{-0.34}$ and  ${\rm log}_{10}  J^k = 19.36^{+0.42}_{-0.41}$ respectively. As is apparent, the \CTA\ sensitivity does not  increase, however, by the same factor. This is due to the fact that the derivation of $ J^k$ in ``ultra-faint'' dSphs is affected by larger statistical errors. Furthermore,  the  sensitivity towards \Seg\ is better than that  in  \UMaII, even though  the peak value of the $J$-factor is smaller. This is again due to the fact that the $J$-profile in \Seg\ is more concentrated and its determination is affected by  smaller statistical errors. Nevertheless, a recent study points out that the mean value quoted in Tab.~\ref{tab:dSphs} is largely overestimated~\cite{Bonnivard:2015xpq}. In view of that, we promote \UMaII\ as the best ``ultra-faint'' dSph  for the study of VHE $\gamma$-rays from  annihilating DM by  \CTA. 

\smallskip
In the upper-right panel of Fig.~\ref{fig:3} we present the bounds in the ($m_{\rm  DM}$,$\sigmav$) plane of a combined analysis from time-limited observations of several dSphs. We stacked four dSphs that could be observed during \CTA\ extragalactic surveys. In particular, we consider the ``classical'' dSphs, \Dra\ and \UMi, and the ``ultra-faint'' ones, \UMaII\ and \Com. The results are presented for the DM DM $\rightarrow b \bar b$ annihilation channel for 50 h of observation on each dSph (dashed lines) and for the cumulative stacked observation of 200 h (solid black line). For comparison, we also show  the sensitivity by  observing   \UMaII\ alone for 200 h (dotted blue line).  As one can see, with a stacking analysis, the \CTA\ sensitivity is     $\langle \sigma v \rangle \simeq 1.0 \times 10^{-23}$ cm$^3$s$^{-1}$ at $m_{\rm DM}\simeq 1$ TeV. This is even weaker than  the projected limit  one can get by observing \UMaII\ alone for just 50 h ($\langle \sigma v \rangle \simeq 8.8 \times 10^{-24}$ cm$^3$s$^{-1}$ at $m_{\rm DM}\simeq 1$ TeV). Indeed,  the large statistical uncertainty  in the determination of the $J$-factor weakens the sensitivity, especially in  ``ultra-faint'' dSphs. As a consequence, if  no improvements are made on the determination of the statistical errors, we suggest to point the IACTs towards the single dSph galaxy with the most promising $J$-factor and the lowest statistical uncertainties (e.g.~\UMi, \UMaII), instead of performing a stacking analysis.

\medskip
The lower-panels of Fig.~\ref{fig:3} shows the sensitivity for 500 h of observation in different annihilation channels for two ``classical''  and one ``ultra-faint''  dSph galaxies. For all dSphs, the strongest sensitivity is obtained for the DM DM $\rightarrow \tau^+ \tau^-$ primary annihilation mode (blue lines) for $m_{\rm DM}\simeq 300$ GeV. In particular, for \Dra\ (left panel) one gets $\langle \sigma v \rangle \simeq 3.1 \times 10^{-24}$ cm$^3$s$^{-1}$, while for the best ``classical'' dSph, \UMi\  (lower-central panel), the
\CTA\ sensitivity reaches $\langle \sigma v \rangle \simeq  1.0 \times 10^{-24}$ cm$^3$s$^{-1}$. In the lower-right panel, we report the sensitivity towards our best ``ultra-faint'' dSph, \UMaII. The bounds are quite strong reaching  $\langle \sigma v \rangle \simeq 7.0 \times 10^{-25}$ cm$^3$s$^{-1}$. For the DM DM $\rightarrow \mu^+ \mu^-$ channel (yellow lines), the sensitivity is weaker and the best limit is $\langle \sigma v \rangle \simeq 2.3 \times 10^{-24}$  cm$^3$s$^{-1}$ for $m_{\rm DM}\simeq 200$ GeV.

\smallskip
For the hadronic channels: DM DM $\rightarrow b \bar b$ (black lines), $t \bar t$ (red lines) and the DM DM $\rightarrow W^+ W^-$ (green lines), the best sensitivity is again achieved towards \UMaII\ for  DM masses around 1 TeV. More specifically, for the DM DM $\rightarrow W^+ W^-$ primary channel the sensitivity  reaches  $\langle \sigma v \rangle \simeq 1.9 \times 10^{-24}$ cm$^3$s$^{-1}$ in the TeV DM mass range. 

This is quite interesting because well-motivated DM models with electroweak interactions (e.g.~supersymmetric Wino DM, Minimal DM, ...) predict DM cross sections into EW bosons of the same order of magnitude, thanks to the large enhancement  coming from the non-perturbative Sommerfeld effect. The derivation of the Sommerfeld enhanced cross sections is dramatically model-dependent (in particular the formation of the resonant peaks of the cross sections as a function of the DM mass as shown in e.g.~Refs.~\cite{Hisano:2006nn, Cirelli:2007xd, Cirelli:2015bda, Garcia-Cely:2015dda}). However, in the non-resonant case, the Sommerfeld enhancement saturates when the de Broglie wavelength of the particle $m_{\rm DM} v$  is of the same order of the mass of the mediator $m_\phi$~(see e.g.~\cite{Cassel:2009wt}). As a consequence, for multi-TeV candidates interacting with the SM particles via the exchange of  heavy mediators (i.e.~heavier than the EW gauge bosons), the Sommerfeld enhanced cross section is independent on the velocity in the usual astrophysical targets used in indirect detection, because it saturates when $\beta \lesssim 10^{-1} \left[m_\phi/(100 \mbox{ GeV}) \right]\,\left[ \left(1 \mbox{ TeV}\right)/m_{\rm DM}\right]$. Our bounds in the ($m_{\rm DM}, \langle \sigma v \rangle$) plane are pretty generic and one can directly compare them with the theoretical predictions of a large class of DM models.  Furthermore, since the Sommerfeld effect can boost the cross sections  into $Z\gamma$ and $\gamma\gamma$ in a relevant way, it is of primary interest to study the $\gamma$-ray line features in the DM spectra towards the dSphs with the most promising $J$-factors (e.g.~\UMi,
\UMaII), with  either the currently operating IACTs or the forthcoming ones like \CTA\ (see Ref.~\cite{Lefranc:2016fgn} for a recent study). 

\begin{figure}[!t]
\centering
\includegraphics[width=.65\textwidth]{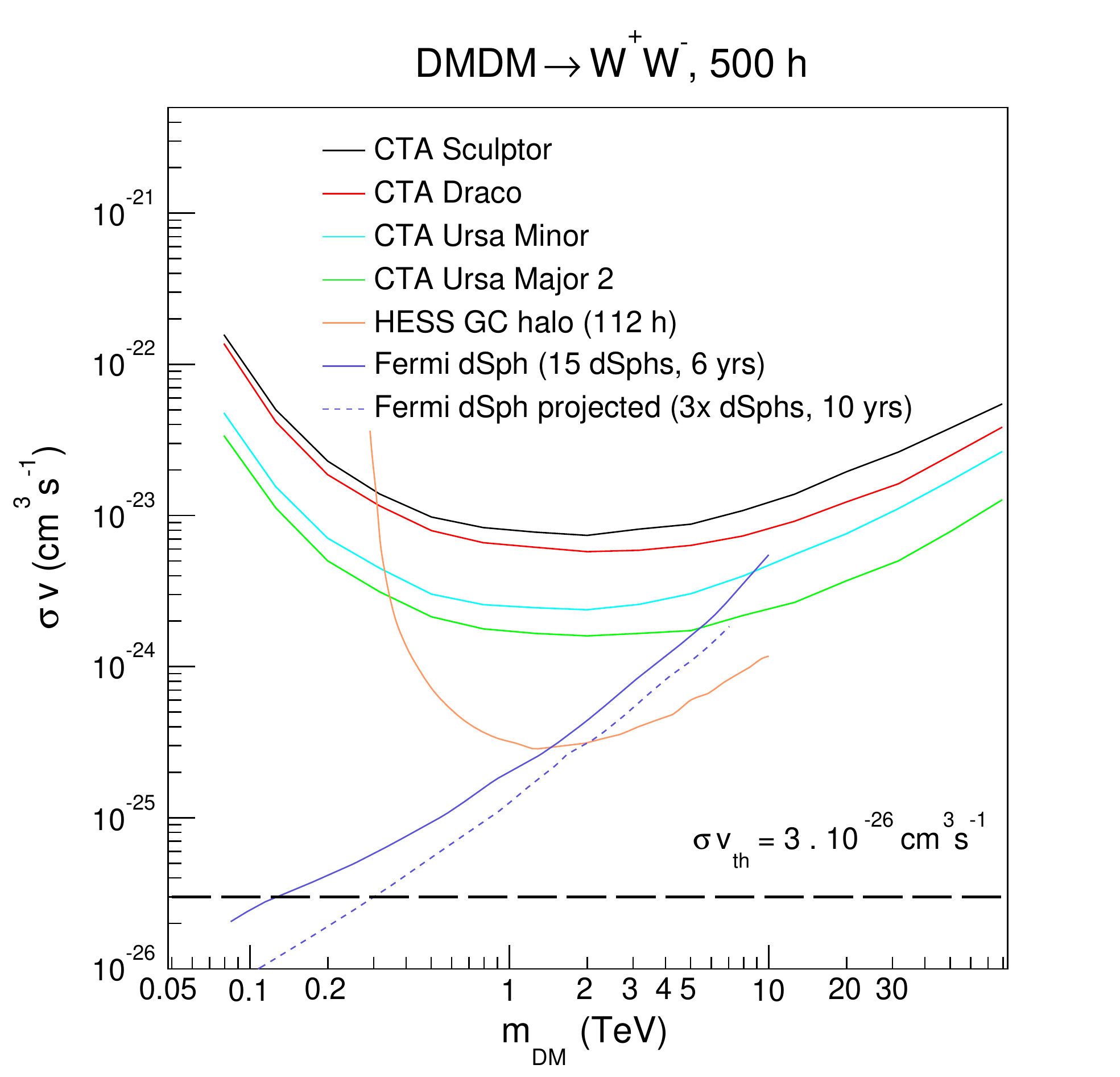}
\caption{\small \CTA\ sensitivity to DM annihilation for the DM DM  $\rightarrow W^+W^-$ channel compared to the most constraining limits to  date. The sensitivities from \CTA\ are obtained by implementing a 2D  statistical approach. They are given by observing  \Scu\ (black line), \Dra\ (red line), \UMi\ (cyan line), \UMaII\ (green line)  for 500 h. We report the \HESS\ bound for 112 h  observations of the GC (orange line) and the limit coming from a stacking analysis on 15 dSph from \FERMI\ (blue line). An estimate of the \FERMI\ sensitivity for 45 dSphs in 10 years of observations is also shown (dashed blue line). For comparison, the value of the thermal cross section  is depicted as the long-dashed black horizontal line.   }
\label{fig:5}
\end{figure}

\medskip
Finally, in Fig.~\ref{fig:5}, we comment on the relative strength of our projected constraints for the DM DM $\rightarrow W^+W^-$ primary mode, with respect to the best limits to date inferred from either other targets or analysis. In particular the red and black  lines are obtained by observing the ``classical'' dSphs \Dra, \Scu\ and \UMi\ for 500 h, while the green line is for pointing the \CTA\ array towards \UMaII\ for the same amount of time. For comparison, we also report the value of the thermal cross section ($\sigmav_{\rm th}=3\times 10^{-26}$ cm$^3$s$^{-1}$) as the long-dashed black line. 

With respect to the  stacking analysis on 15 dSph observations from \FERMI~\cite{Ackermann:2015zua} (solid blue line), our bounds become competitive for DM masses above  several  TeV. This remains the case in the optimistic scenario of 10-year observations with \FERMI~\cite{Anderson14} of 45 dSph galaxies (dashed blue line). Hence,  by observing dSph galaxies, \FERMI\ and \CTA\ will be able to probe the properties of a large class of DM models. Indeed, on one hand, \FERMI\ will be the leading experiment for probing relatively light  DM candidates ($m_{\rm DM}\lesssim 500$ GeV) with the thermal cross section, while on the other hand, \CTA\ will be the major experiment for the study of well-motivated heavy WIMPs  ($m_{\rm DM}$ above few TeV) with Sommerfeld enhanced cross sections into electroweak bosons. 

With respect to the \HESS\ bounds obtained by observing the GC for 112 h~\cite{Abramowski:2011hc} (solid orange line), the constraints derived here cover a broader range of DM masses (from 50 GeV to 80 TeV). More specifically, our projected  sensitivity towards \UMaII\ overtakes the current \HESS\ sensitivity  in the GC for $m_{\rm DM}$ below roughly 300 GeV and above tens of TeV. Nevertheless, it is worth stressing that the IACTs sensitivity to DM annihilations in the GC can be largely degraded  by the poor knowledge of the Milky Way's DM profile in the inner few kpc. Indeed, as already pointed out in Sec.~\ref{Sec:CTA}, the ON-OFF method in the GC center is only performant for a cuspy DM profile. As a consequence, since the constraints coming from our list of dSph galaxies are affected by much smaller uncertainties, they are in principle more robust and stable.

\section{Summary}
\label{Sec:Summary}
Dwarf Spheroidal Galaxies of the Milky Way are probably the cleanest laboratory for looking at VHE $\gamma$-rays from DM annihilations or decays, due to their proximity, high DM content, and low stellar and other non-DM emission foregrounds. In this work, we study the future \CTA\ sensitivity to DM annihilations towards dSph galaxies, considering a broad range of DM masses (from 50 GeV up to 80 TeV), and  several primary channels (DM DM $\rightarrow$ $\mu^+\mu^-$, $\tau^+\tau^-$, $b\bar{b}$, $t\bar{t}$, and $W^+W^-$).  In particular, we derive our bounds by pointing the \CTA\ array towards 5 high luminous ``classical'' dSph galaxies (\Car, \Dra, \For, \Scu, \UMi) and 5 ``ultra-faint'' ones (\Boo, \Com, \Seg, \UMaI, \UMaII). 
We also present the \CTA\ sensitivity from a combined analysis of 4 dSph galaxies (\Dra, \UMi, \UMaII\ and \Com). 

We assess the \CTA\ sensitivity by: $i)$ using, for each dSph galaxy, a recent determination of the $J$-factor and its statistical error coming from careful studies of  stellar-kinematic data (see e.g.~\cite{Geringer-Sameth:2014yza, Bonnivard:2015xpq, Bonnivard:2015tta}); $ii)$ considering the most up-to-date CR background determination  from a full \CTA\ Monte Carlo simulation; $iii)$ applying three different statistical approaches in order to show how the \CTA\ sensitivity gets improved by taking full advantage of the peculiar spectral and spatial features  of the DM signal with respect to the irreducible background; and $iv)$ including a Gaussian term in the likelihood that measures the impact of the $J$-factor statistical uncertainties on the \CTA\ sensitivity. 

\smallskip
We demonstrated that is  crucial to consider the dSph galaxies as extended objects in terms of $\gamma$-rays observations by \CTA. We  found that a spectral analysis  improves the \CTA\ sensitivity (from 20\% to 40\%), especially for primary channels with peculiar spectral features (e.g.~DM DM $\rightarrow W^+W^-$). We also found that the \CTA\ sensitivity can be further improved by implementing a 2D statistical approach   that, in addition to the spectral features of the DM signal, takes  advantage of the spatial extent of a given dSph as well. Indeed, instead of considering a single RoI  that covers the entire angular size of a given dSph, a spatial analysis allows to integrate more signal than background by decreasing the angular extents of the RoIs. This is particularly relevant for dSph galaxies with steep $J$-profiles (\UMi\ and \Com).

\smallskip
We found, from the $J$-factors in Ref.~\cite{Geringer-Sameth:2014yza} and our statistical analysis, that the best dSphs for studying VHE $\gamma$-rays from DM annihilations are  \UMi\ (best ``classical'' dSph) and \UMaII\ (best ``ultra-faint'' dSph).   For all dSphs, the strongest sensitivity is achieved for the DM DM $\rightarrow \tau^+ \tau^-$ primary annihilation mode at $m_{\rm DM}\simeq 300$ GeV. In particular, with 500 h of observation towards \UMaII, one can rule out annihilation cross section larger than $7.0 \times 10^{-25}$ cm$^3$s$^{-1}$. 

For the hadronic channels and the DM DM $\rightarrow W^+W^-$ mode, the best sensitivity is again achieved towards \UMaII\ at $m_{\rm DM}\simeq 1$ TeV. More specifically, the best sensitivity is obtained for the DM DM $\rightarrow W^+W^-$ primary mode ($\sigmav \simeq 1.9\times 10^{-24}$ cm$^3$s$^{-1}$ in the multi-TeV mass range). This is very interesting, because well-motivated heavy DM candidates with electroweak interactions possess   annihilation cross sections towards dSph galaxies of the same order of magnitude, thanks to the non-perturbative Sommerfeld effect that can  significantly boost the DM annihilations into all electroweak bosons.

\section*{Acknowledgments} 
The authors would like to thank Emmanuel Moulin and Joe Silk for fruitful discussions during the elaboration of this work. P.P. acknowledge the support of the European Research Council project 267117 hosted by Universit\'e Pierre et Marie Curie-Paris 6, PI J.~Silk. 

\bibliographystyle{My}
\small
\bibliography{bibl}

\end{document}